\documentclass[lettersize,journal]{IEEEtran}
\usepackage{amsmath,amsfonts}
\usepackage{algorithmicx}
\usepackage{algorithm}
\usepackage{array}
\usepackage[caption=false,font=normalsize,labelfont=sf,textfont=sf]{subfig}
\usepackage{textcomp}
\usepackage{stfloats}
\usepackage{url}
\usepackage{verbatim}
\usepackage{graphicx}
\usepackage{braket}
\usepackage{algpseudocode}
\usepackage{float}

\makeatletter
\let\NAT@parse\undefined
\makeatother
\usepackage{hyperref}  
\usepackage[numbers,sort&compress]{natbib}

\begin{document}

\title{Auxiliary Task-based Deep Reinforcement Learning for Quantum Control}

\author{Shumin Zhou, Hailan Ma, Sen Kuang{*},
and Daoyi Dong
\thanks{This work was supported by
the Australian Research Council’s Future
Fellowship funding scheme under Project FT220100656, and U.S. Office of
Naval Research Global under Grant N62909-19-1-2129.
\textit{(*Corresponding Author: Sen Kuang)}
}
\thanks{S. Zhou and S. Kuang are with the Department of Automation, University of Science and Technology of China, Hefei 230027, China (e-mail: shuminzhou@mail.ustc.edu.cn, skuang@ustc.edu.cn).}
\thanks{H. Ma and D. Dong are with the School of Engineering and Information Technology, University of New South Wales, Canberra ACT 2600, Australia (e-mail: hailanma0413@gmail.com, daoyidong@gmail.com).}
}



\maketitle

\begin{abstract}
Due to its property of not requiring prior knowledge of the environment, reinforcement learning has significant potential for quantum control problems. In this work, we investigate the effectiveness of continuous control policies based on deep deterministic policy gradient. To solve the sparse reward signal in quantum learning control problems, we propose  an auxiliary task-based deep reinforcement learning (AT-DRL) for quantum control. In particular, we first design a guided reward function based on the fidelity of quantum states that enables incremental fidelity improvement. Then, we introduce the concept of an auxiliary task whose network shares parameters with the main network to predict the reward provided by the environment (called the main task). The auxiliary task learns synchronously with the main task, allowing one to select the most relevant features of the environment, thus aiding the agent in comprehending how to achieve the desired state. The numerical simulations demonstrate that the proposed AT-DRL can provide a solution to the sparse reward in quantum systems, and has great potential in designing control pulses that achieve efficient quantum state preparation.
\end{abstract}



\section{Introduction}
\IEEEPARstart{M}anipulation of quantum systems plays a significant role in many fields such as quantum computing \cite{1,2,3}, quantum high-precision measurements \cite{4,5,6}, chemical physics \cite{dong2022quantum} and quantum communication networks \cite{8}. To achieve quantum operations with high efficiency, various control approaches \cite{dong2022quantum}, such as optimal control \cite{11,12,13}, sliding mode control \cite{15}, robust and learning control \cite{16,17,wu2019learning} and Lyapunov control \cite{18,19,20,21} have been proposed to solve this problem. Traditional methods may rely on the model of systems and thus have limited applications for complex quantum systems \cite{dong2010quantum}.

Reinforcement learning (RL) \cite{kaelbling1996reinforcement}, a branch of machine learning in artificial intelligence \cite{22}, has proven to be a powerful tool to solve a wide range of complex problems, such as Alphago and Atari. Owing to its potential in searching for control policy without prior knowledge of the environment \cite{31}, RL is naturally introduced to search for the optimal control fields that maximize a desired performance \cite{25}. In particular, Q-tables have been introduced to solve the shortest path problem \cite{26} and policy gradient methods have been used for approximating compilations of quantum unitary transformations \cite{27}. In recent decades, the introduction of deep learning to the RL framework has greatly enhanced its control performance on high dimensional state (action) spaces and has been  introduced to various quantum tasks \cite{shindi2023model}. For example, the control transport of a quantum state by adiabatic passage through an array of semiconductor quantum dots has been realized by deep Q network (DQN) that utilizes a neural network to predict its Q-value function rather than a Q-table \cite{28}. Furthermore, DQN has also demonstrated its capability of realizing multi-qubit gates with high efficiency and precision \cite{29} and estimating quantum system parameters with improved precision \cite{30}.

In practice, the effectiveness of quantum control relies heavily on the accuracy of control laws. However, the existing algorithms such as DQN can only output partially feasible actions. This results in an insufficient fidelity of the target state and the network converges more slowly or even cannot converge at all. The deep deterministic policy gradient (DDPG) algorithm is an extension of DQN aiming at dealing with continuous control problems \cite{32}. It not only absorbs the advantages of single-step updating of policy gradients in Actor-Critic but also the skill of DQN for Q-value estimation. Such an Actor-Critic framework allows it to explore continuous control fields with more efficiency. Now, DDPG has been recognized as an important tool for solving complicated problems and achieving efficient control for quantum systems. For instance, the deep deterministic policy gradients techniques have been integrated to optimize the fidelity of the superconducting quantum circuit \cite{33}. A general framework of the quantum-inspired DDPG method has been invented to efficiently solve both classical and quantum sequential decision problems (e.g., state preparation) in the continuous domain \cite{34}.

The goal of RL is to derive an agent’s policy that maximizes its utility in sequential decision-making problems. In the standard formulation, the utility of an agent is defined via its reward function, which plays a critical role in enhancing the agent’s learning process  \cite{44}. When applying RL to quantum systems, its control effects are greatly limited by the sparse reward since a random search makes it difficult for the quantum system to evolve into the final state with dense reward signals, thus providing limited information for neural networks to suggest good actions \cite{31}. To realize an efficient control of quantum systems, we utilize quantum fidelity (which is widely used in the quantum information community) to design an enhanced reward function \cite{31,34}. Inspired by the potential energy reward function \cite{47}, we design a guided reward function that encourages the RL agent to find a good policy that gradually increases the fidelity and prevents the premature termination of exploration.


Recently, the auxiliary task setting has been introduced to boost the performance of the original RL (defined as the main task). For example, two auxiliary tasks involving depth map prediction and closed-loop detection in the navigation enable the agent to explore the environment with more efficiency by detecting whether the current location has been reached even when visual cues are scarce in I-maze \cite{49}. An auxiliary task of direction prediction accelerates the learning process by providing extra gradients \cite{50}. In the 3D map visual navigation task, a reward prediction auxiliary task has been added to the Asynchronous Advantage Actor-Critic (A3C) algorithm, which makes the agent significantly outperform the previous state-of-the-art on Atari \cite{51}. To efficiently use the reward signals from quantum systems, we introduce an auxiliary task aiming at predicting new reward signals before the RL agents obtain the actual reward from the environment. The auxiliary task learns synchronously with the main task, allowing the task to select the most relevant features of the environment that are key to suggesting good actions towards the desired states \cite{52}.



 To effectively control quantum systems, we propose an auxiliary task-based deep reinforcement learning (AT-DRL) for quantum control. In particular, we utilize the deep deterministic policy gradient (DDPG) method to generate continuous control fields for quantum systems. In addition, we design a guided reward function based on the fidelity of quantum states and utilize an auxiliary task with a neural network to fit the reward signal, to make the best use of the reward information in quantum systems. The main task is closely correlated with the auxiliary task and the shared parameters settings allow the main network to capture useful features that provide important implications for suggesting good actions. To demonstrate the effectiveness of the proposed method, numerical simulations of state preparations are presented and compared.

 The main contributions of this paper are summarized as follows.
\begin{enumerate}
    \item To overcome the sparse reward signal for quantum control problems, we design a guided reward function to incentivize the RL agent to incrementally improve its performance towards high fidelity.
    \item Apart from the main task, an auxiliary task is introduced to predict the reward given by the environment. By sharing the parameters between the two tasks, the RL agent is able to select the most relevant features of the environment that are key to suggesting good actions toward the desired states.
    \item To verify the effectiveness of the proposed method, numerical simulations on one-qubit, two-qubit, and eight-qubit quantum systems are implemented, respectively. The supervisory of AT-DRL over two traditional DRL methods (DQN and DDPG) demonstrates that the proposed method can achieve efficient state preparation even with sparse reward signals.
\end{enumerate}

The rest of this paper is organized as follows. Section II introduces several basic concepts about RL, DDPG, and quantum dynamics. The design and implementation of the AT-DRL algorithm are described in detail in Section III. In Section IV, numerical results for several quantum systems are presented. Concluding remarks are presented in Section V.


\section{Preliminaries}
\subsection{Markov Decision Process and Deep Deterministic Policy Gradient Policy}
Markov decision process is a mathematical model of a sequential decision-making process that is involved in reinforcement learning. A Markov decision process consists of a five-tuple $\left[ \textit{S, A, P, R, $\gamma$} \right]$ \cite{37}, where $\textit{S}$ denotes the state space, $\textit{A}$ represents the action space, and \textit{P} $ \in \left[0,1\right)$ is the state transfer probability.
 \textit{R: S $\times$ A $\rightarrow$ $\mathbb{R}$} is the reward function, and $\gamma \in \left[0,1\right]$ is the discount factor.
 At each time step $t \in \left[0,\mathrm{T}\right]$, the state of the agent is $s_t \in S$. The action for the agent is chosen according to the policy $\pi$: $a_t = \pi(s_t)$. After performing the action $a_t$, the agent transfers to the next state $s_{t+1}$ and receives the scalar reward signal $R_t$ from the environment. RL aims to determine the optimal action $a_t$ for each state $s_t$ to maximize the cumulative future discounted return $R_t=\sum_{k=0}^{T-t} \gamma^{k} r_{t+k}$.

Due to the integration of deep learning and reinforcement learning, DQN was proposed as a powerful deep reinforcement learning (DRL) method and applied to address discrete problems
 \cite{mnih2013playing,mnih2015human}. This idea was extended to the continuous problem, which was formulated as Deep Deterministic Policy Gradient Policy (DDPG). Proposed by the DeepMind team \cite{38,39}, DDPG is really helpful when the dimension of the action space is high, where DQN fails to learn a good policy. The DDPG algorithm is actually a classical Actor-Critic algorithm consisting of an Actor network and a Critic network. In this architecture, the Actor network generates actions while the Critic network evaluates these actions based on the current value function \cite{konda1999actor}. By utilizing these two networks, reinforcement learning is capable of successfully finding a good policy for continuous control. In DDPG, both the policy-based (also known as Actor) network and the value-based (also known as Critic) network are divided into two parts: a current network for updating at each step and a target network for computing the predicted Q values and actions. The target network copies the parameters of the current network in a regular period to obtain a relatively stable Q value that is helpful for the convergence of the model.

To train the neural networks, experience replay is utilized \cite{mnih2013playing}, where experiences or transition $(s_t,a_t, r_t, s_{t+1}, done)$ are stored in a large memory pool and are sampled from the pool to update the parameters of the neural networks. This helps to achieve independent identical distribution and thus contributes to accelerated convergence \cite{39}.

Exploration is essential for agents, yet a set of deterministic networks can only output a deterministic action for an input state. To solve this, noise can be added to the output actions to improve the exploration  ability of agents. For example, the Ornstein-Uhlenbeck process \cite{41} is a common choice to introduce noise to the action space. It can be defined by the following stochastic differential equation (in the one-dimensional case):
\begin{equation}\label{eq1}
dN_t = \theta(\mu - N_t)dt + \sigma dB_t,
\end{equation}
where $dN_t$ denotes the value inscribed and $\mu$ denotes its weight. $\theta$ determines the speed of mean reversion and is an important feature that distinguishes one Ornstein-Uhlenbeck process from others. Large $\theta$ leads to smaller system disturbance, suggesting that the current value is near the mean value. $B_t$ represents the standard Brownian motion that acts as an external random noise with $\sigma \textgreater 0$ denoting its weight. When the initial perturbation is a single-point distribution at the origin (i.e., limited to $N_0 = 0$) and $\mu = 0$, the solution of the above equation is
\begin{equation}\label{eq2}
N_t = \sigma \int_{0}^{t} e^{\theta(\tau-t)} dB_t.
\end{equation}

\subsection{Quantum Dynamics}
The state of a classical bit is either 0 or 1, whereas a state of one-qubit systems can be described as a linear combination of the ground states
$\ket{0}$ and $\ket{1}$, which is also called the superposition state
\begin{equation}\label{eq3}
\ket{\psi} = \alpha\ket{0} + \beta\ket{1},
\end{equation}
where $\alpha$ and $\beta$ are complex numbers that satisfy the normalization condition $|\alpha|^2 + |\beta|^2 = 1$. This applies to multiple qubits, for example, the state of two-qubit systems is written as
\begin{equation}\label{eq4}
\ket{\psi} = \alpha_{00}\ket{00} + \alpha_{01}\ket{01} + \alpha_{10}\ket{10} + \alpha_{11}\ket{11},
\end{equation}
where $\ket{00}$, $\ket{01}$, $\ket{10}$, and $\ket{11}$ are four computational states and  $\alpha_{jk}$ are complex numbers that satisfy normalization. For multiple qubits, when the states cannot be written as the product of two states in the subsystems (i.e., $\ket{\psi_1}\ket{\psi_2}$), we call them entangled quantum states. For example, Bell states are the most widely used two-qubit entangled states, formulated as

\begin{equation}\nonumber
\begin{aligned}
\ket{\phi^+} =(\ket{00}+\ket{11})/\sqrt{2},\\
\ket{\phi^-} = (\ket{00}-\ket{11})/\sqrt{2},\\
\ket{\psi^+} =(\ket{01}+\ket{10})/\sqrt{2},\\
\ket{\psi^-} =(\ket{01}-\ket{10})/\sqrt{2}.
\end{aligned}
\end{equation}

Denote $\ket{\psi(t)}$ as the state vector for a closed quantum system, its evolution can be described by the Schr\"{o}dinger equation \cite{42}:
\begin{equation}\label{eq5}
\begin{aligned}
    i\hbar\frac{\partial}{\partial{t}}\ket{\psi(t)}=[H_0+\sum_k u_k H_k]\ket{\psi(t)}, \ket{\psi(0)} = \ket{\psi_0},
\end{aligned}
\end{equation}
where $\hbar$ is the reduced Planck constant and is usually set as 1 in the atomic unit system. $H_0$ denotes the free Hamitonian,  $\sum_k u_k H_k$ represents the control Hamitonian with $u_k$ being the external control fields. The solution of (5) can be given as equation (6) \cite{43}, where $U(t,t_0)$ is an evolution operator $U$ satisfying the following equation (7):
\begin{equation}\label{eq6}
\begin{aligned}
\ket{\psi(t)}=U(t,t_0)\ket{\psi(t_0)},
\end{aligned}
\end{equation}
\begin{equation}\label{eq7}
\begin{aligned}
    i\hbar\frac{\partial{U}}{\partial{t}}=\widehat{H}U,
\end{aligned}
\end{equation}
where $\widehat{H}$ donates a Hamiltonian operator that represents the total energy of a wave function.

\section{METHODOLOGY}

In this section, the framework of auxiliary task-based deep reinforcement learning for quantum control is presented. Then two key elements that contribute to the improved control performance for quantum systems are introduced, including guided reward function design and reward prediction using an auxiliary task. Finally, the implementation detail of AT-DRL for quantum control is summarized.

\subsection{Framework of Auxiliary Task-based Deep Reinforcement Learning for Quantum Control}

\begin{figure*}[!t]
\centering
\includegraphics[width=1\linewidth]{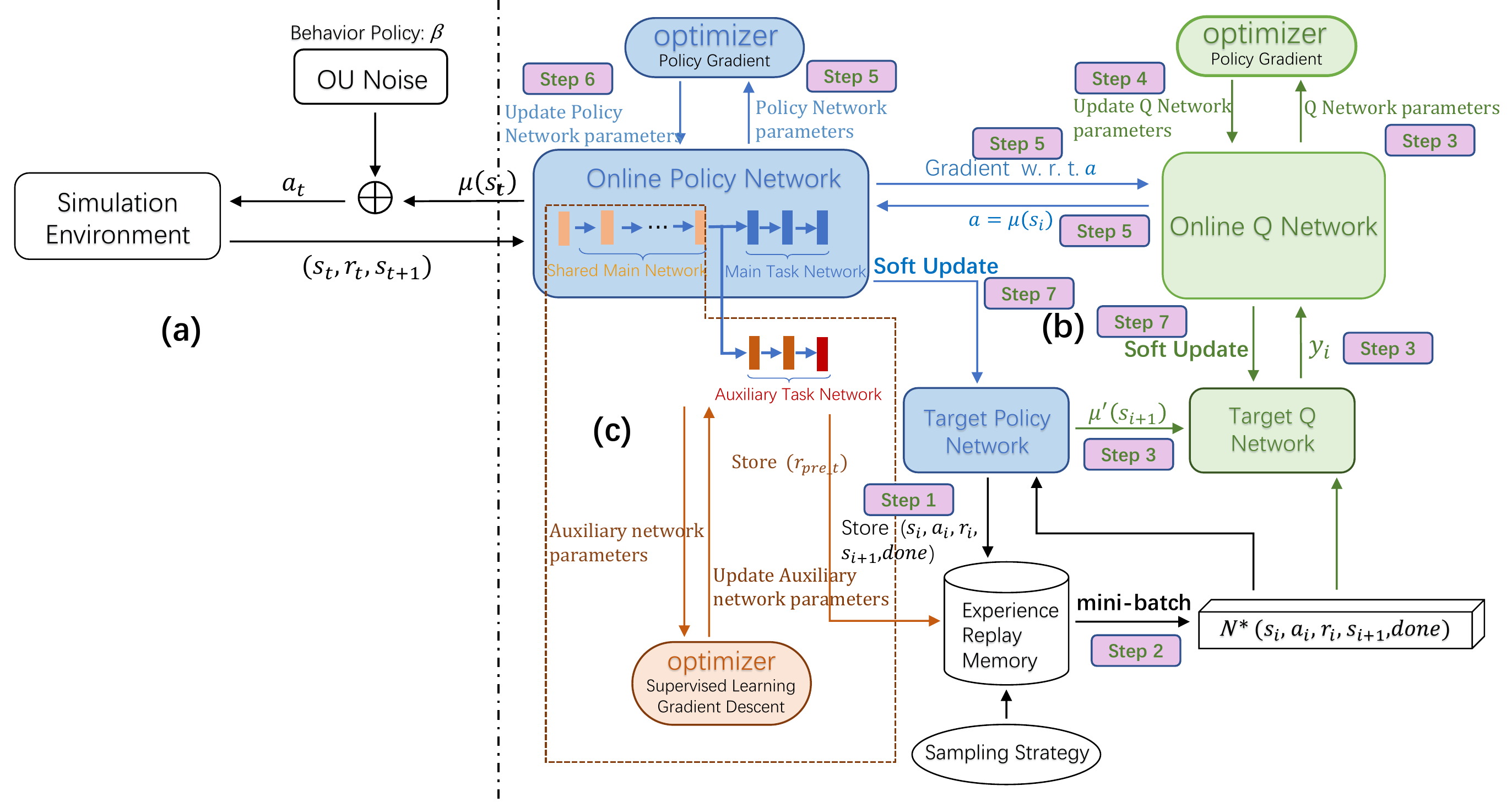}
\caption{The framework of AT-DRL for quantum control. (a) Interaction of the agent with the environment. The agent needs to constantly interact with the environment in order to enhance their understanding. The actor selects an action $a_t$ based on the behavior policy and transmits it to the simulation environment for execution. Upon completing the execution of $a_t$, the simulation environment provides a reward $r_t$ and a new state $s_{t+1}$. The specific design format of the reward $r_t$ will be elaborated in Section III-B. (b) The actor stores the state transition process including $(s_i,a_i,r_i,s_{i+1},done)$ in the replay memory buffer, thereby building a training dataset for the online network. (Step 1) Subsequently, a mini-batch of $N$ transition data is randomly sampled from the replay memory buffer, serving as the training data for the online policy network and the online Q network. The parameters of Actor network and Critic network are optimized using the Adam optimizer (Step 2). (c) Reward Prediction Auxiliary Task. An auxiliary neural network is established to predict rewards provided by the environment, which mitigates the issue of reward sparsity. The output $r_{pre\_t}$ of the auxiliary network is also added to the replay memory buffer. The network shares some parameters with the main network, which facilitates the iteration of the main task network during auxiliary task training.}
\label{fig:framework}
\end{figure*}

To achieve accurate control of quantum systems, we utilize DDPG as the basic algorithm owing to its capability of generating continuous control policies. As an extension of DQN, DDPG contains the Actor-Critic framework \cite{konda1999actor}, where the Actor is used for generating policies $\pi(a|s)$ and output actions for state $s$, and the Critic is used to evaluate the effectiveness of a strategy using a value function $Q(s,a)$. Both networks have their corresponding target networks. Thus, four networks in all are involved in the DDPG framework, i.e., Actor-network $\mu(\cdot|\theta^{\mu})$, the Critic network $Q(\cdot|\theta^{Q})$, the Target Actor network $\mu'(\cdot|\theta^{\mu'})$, and the Target Critic network $Q'(\cdot|\theta^{Q'})$. The process of searching for a good policy using DDPG is defined as the main task in this work.


In the standard RL paradigm, the training of RL agents is realized through trial and error, where the environment provides a scalar reward that encourages good actions and weakens bad actions. When applying it to control tasks, it is highly desirable to design a suitable reward function to overcome the sparse reward signal, which is usually the case in quantum control problems \cite{31}. Inspired by the concept of the potential energy reward function \cite{47}, we design a guided reward function that encourages the agent to find a good policy for achieving incremental improvements in fidelity. To make full use of the valuable reward, we introduce an auxiliary task aiming at predicting new reward signals for new states. To associate the main task and the auxiliary task together, the parameters of the auxiliary task are partially shared  with the Actor in the main task and are optimized in a sparse way \cite{52}.

The whole framework of AT-DRL for quantum control is illustrated in Fig. \ref{fig:framework}, with three parts. (a) The agent interacts with the environment to enhance their understanding, where the actor selects an action $a_t$ based on the behavior policy and transmits it to the simulation environment for execution. Upon completing the execution of $a_t$, the RL agent evolves into a new state $s_{t+1}$. (b) The guided reward signal $r_t$ is obtained based on the fidelity between the current state $s_{t+1}$ and the target state. The state transition process including $(s_i,a_i,r_i,s_{i+1},done)$ in the replay memory buffer and a mini-batch of transitions is randomly sampled from the replay memory buffer, serving as the training data for the online policy network and the online Q network. (c) An auxiliary neural network is established to predict rewards provided by the environment, which mitigates the issue of reward sparsity. The output $r_{pre\_t}$ of the auxiliary network is also added to the replay memory buffer. The network shares some parameters with the main network, which facilitates the iteration of the main task network during auxiliary task training.

\subsection{Guided Reward Function Design}
The goal of reinforcement learning (RL) is to derive an agent's policy that maximizes its utility in sequential decision-making tasks. In the standard formulation, the utility of an agent is defined via its reward function, which plays a critical role in providing sound specifications of the task goals and supporting the agent's learning process \cite{44}. There are different perspectives on reward design, which differ in the investigation, which is extremely important in reinforcement learning.

In quantum control, fidelity is widely used to measure the similarity between quantum states and therefore evaluates the performance of the control task \cite{nielsen2010quantum}. For a state transfer problem, we can utilize the following fidelity
\begin{equation}\label{eq14}
\begin{aligned}
    F(\langle\psi_f|\psi\rangle) = |\langle\psi_f|\psi\rangle|^2,
\end{aligned}
\end{equation}
where $\ket{\psi}$, and $\ket{\psi_f}$ denote the actual state and the target state, respectively. For each step $t$ with the quantum state as $\ket{\psi_t}$, we have $F_t=|\langle\psi_f|\psi_t\rangle|^2$.

In order to drive the quantum system to the target state, we design a binary ``guided'' reward function. First, the reward function is divided into two phases based on whether the ideal fidelity has been achieved. Then, to prevent premature termination of exploration, we draw on the concept of a potential energy reward function \cite{47} to incentivize further improvement in fidelity. Inspired by the concept of power in physics, the reward function assigns a positive reward to the agent when it transitions from a high potential energy state to a low potential energy state, and a negative reward is given when the agent moves from a low potential energy state to a high potential energy state. This approach follows the paradigm of the potential Markov Decision Process (MDP), whose reward function is formulated as
 \begin{equation}
\begin{aligned}
& R^{\prime}\left(s, a, s^{\prime}\right)=R\left(s, a, s^{\prime}\right)+P\left(s^{\prime}\right) \\
& P\left(s^{\prime}\right)=\Phi\left(s^{\prime}\right)-\gamma \Phi(s)
\end{aligned}
\end{equation}
where $R$ is the original reward function and $R^{'}$ is the modified reward function. $P$ denotes the potential energy function, and $\Phi$ denotes potential energy. By incorporating the potential energy difference between the states as an extra reward $P$, the optimal solution of the MDP remains unchanged.

For quantum problems, we define the fidelity gap between the current state and the target state as $e_t=1-F_t$ at the step moment $t$ and utilize it as the potential energy to design a guided reward function. Here, the ``guided'' means that the quantum system is evolving toward the target state when $e_t < e_{t-1}$, allowing it to acquire the reward from the environment. In particular, we utilize the following function

\begin{equation}\label{eq15}
r_{t}\! =\!\left\{\begin{array}{c}
10000+1000 *\left(e_{t-1}-e_t\right), F_t \geq f_0 \\
F_t * 100+1000 *\left(e_{t-1}-e_{t}\right), F_t<f_0 \text { or $t$ } \geq N_{max}
\end{array}\right.
\end{equation}
where $F_t$ is the fidelity, $f_0$ is the ideal fidelity, $t$ denotes the number of exploration steps.

The reward signal is given using two phases based on the criterion of realizing $f_0$ \cite{45}, where the first involves a fixed reward of 10000 and the second involves a reward proportional to the gap between the current and target states. When fidelity reaches the ideal level, the fixed reward aims to maintain fidelity above $f_0$. If the current fidelity falls below $f_0$, the reward is directly proportional to the fidelity, facilitating the efficient evolution of the system toward the desired state while assigning smaller rewards for low fidelity levels \cite{46}.

\subsection{Predicting Reward using Auxiliary Task}
During the evolutionary process, the quantum system must identify states with high rewards to effectively learn value functions and strategies. However, the rewards given by the environment tend to be sparsely distributed and might be given at the end of control pulses \cite{48}. For a simple problem with a limited state space, it is easy for the agent to randomly explore, and effective state transfers at a certain percentage can be guaranteed. As the problem's complexity increases, the probability of exploring in a random way becomes small. The sparse feedback signals fail to indicate optimal exploration directions for the agent, making it difficult to develop local knowledge. The blind exploration results in a low sample efficiency in experience replay and therefore low efficiency for the state preparation problem. In this case, the RL algorithm struggles to converge. In the evolutionary process, what and how should the agent learn in the case where the reward is not immediately known?

To overcome the issue of sparse rewards, an auxiliary task is proposed to predict the rewards with the purpose of making the rewards dense enough to provide the agent with strong guidance. In fact, the introduction of an auxiliary task enables the RL agent to explore the environment more efficiently \cite{49,50,51} and has constituted a novel approach whereby reinforcement learning-based navigation algorithms can be leveraged for real-world street view navigation \cite{mirowski2018efficient}. Considering the sparse reward obtained from fidelity, we propose an auxiliary task-based DRL for quantum control. Specifically, we incorporate a neural network to fit the environment's reward signal, allowing the agent to predict the reward signal of executing an action based on the output of the auxiliary task neural network before the execution ends. To update the parameters of the main network and the auxiliary networks efficiently, we utilize sparse sharing \cite{52}.

Sparse sharing means that each task corresponds to a network and all networks perform their respective tasks independently but share some network parameters. When training auxiliary tasks, parameter updates for the shared part facilitate network iteration of the main task. By sharing these parameters, the agent must balance improving its performance with respect to the global reward $r_t$ with improving performance on the auxiliary tasks. If the main task and the auxiliary task are strongly correlated, then these two networks have a high parameter overlap rate. Conversely, if they are weakly correlated, then their parameter overlap rate is low. Although each task is only trained with its corresponding sub-network, a part of the parameters is shared by both tasks at the same time. Therefore, this shared part is updated multiple times in the learning process of every network.

To predict the reward with high efficiency, the parameters of the auxiliary network are optimized using the gradient descent method of supervised learning. At the time $t$, the main network outputs the action $a_t$, which is then executed by the agent. The agent receives a reward $r_t^*$ from the environment, which is also the supervisory signal for the auxiliary task. The output of the auxiliary network is the predicted reward $r_t$. To make the output value approach the actual reward given by the environment as possible, we can update the parameters of the auxiliary network using the following squared loss function, given by equation (16):

\begin{equation}\label{eq16}
\begin{aligned}
    J(\theta)=(r_t^* - r_t)^2.
\end{aligned}
\end{equation}

\subsection{Implementation}

The idea of the DDPG algorithm for the quantum state preparation problem is as follows. First, the network parameters $\theta^{Q}$, $\theta^{\mu}$, $\theta^{Q'}$, $\theta^{\mu'}$ and the experience pool Replay Buffer of the four networks are initialized, as well as the initial state $s_{t_0}$ of the quantum system. In this paper, we use piecewise constant pulses for control and set the duration of each segment to be equal. The total time $T$ is divided into $N$ equal time steps with each step size being  $dt=T/N$. At each time step, the action $a_t$ is determined by the current policy and the action noise. After each $d_t$, depending on the current state, the environment rewards the system with $r_t$ and observes the next state $s_{t+1}$. Each round stores the quintet $(s_t,a_t,r_t,s_{t+1},done)$  in the Replay Buffer. At the end of the exploration process, a specified number of quintets $(s_i,a_i,r_i,s_{i+1},done)$ are randomly selected from the Replay Buffer for the network to learn. The Critic network updates its parameters by continuously decreasing the loss value, i.e.,
\begin{equation}\label{eq8}
\begin{aligned}
    y_i = r_i + \gamma Q'(s_{i+1},\mu'(s_{i+1}|\theta^{\mu'})|\theta^{Q'}),
\end{aligned}
\end{equation}
\begin{equation}\label{eq9}
\begin{aligned}
    L = \frac{1}{N} \sum_i (y_i - Q(s_i,a_i|\theta^Q))^2,
\end{aligned}
\end{equation}
where $Q(\cdot|\theta^{Q})$ denotes the output of the Critic network, $\mu'(\cdot|\theta^{\mu'})$ is the output of the Target Actor network, and $Q'(\cdot|\theta^{Q'})$ stands for the output of the Target Critic network. Parameters $s_i$, $a_i$, and $r_i$ represent the state, action, and reward value at the $i$th moment, respectively.

The policy network uses the following gradient descent method for parameter updates:
\begin{equation}\label{eq10}
\begin{aligned}
\nabla_{\theta^\mu} J \approx \frac{1}{N} \sum_i \nabla_a Q\left(s, a \mid \theta^Q\right)|_{s=s_i, a=\mu\left(s_i\right)} \nabla_{\theta^\mu} \mu\left(s \mid \theta^\mu\right)|_{s_i}.
\end{aligned}
\end{equation}

The target network update equation is
\begin{equation}\label{eq11}
\begin{aligned}
\theta^{Q'} \leftarrow \tau \theta^Q + (1-\tau)\theta^{Q'},
\end{aligned}
\end{equation}
\begin{equation}\label{eq12}
\begin{aligned}
\theta^{\mu'} \leftarrow \tau \theta^{\mu} + (1-\tau)\theta^{\mu'},
\end{aligned}
\end{equation}
where $\mu(\cdot|\theta^{\mu})$ denotes the output of the Actor network, $\theta^{\mu}$, $\theta^{\mu'}$, $\theta^{Q}$, and $\theta^{Q'}$ represent the parameters of the Actor network, Target Actor network, Critic network, and Target Critic network, respectively.

After every 100 episodes, we output the evaluation's average reward and fidelity to check the degree of network learning. When a stable state is reached, the network is thought to have learned good parameters. The complete execution process of the neural network is demonstrated in Algorithm 1.
\begin{algorithm*}[t]
\caption{Algorithm description for AT-DRL}
\hspace*{0.02in} {\bf Input:}
maximum steps $N_{max}$, number of quantum bits $n$, state space dimension $n_{features}=2*n$, action space dimension $n_{actions}$ \\
\hspace*{0.02in} {\bf Output:}
optimal control values
\begin{algorithmic}[1]
\State Randomly initialize the Critic network $Q(s,a|\theta^{Q})$ and the Actor network $\mu(s|\theta^{\mu})$ with weights $\theta^{Q}$ and $\theta^{\mu}$;
\State Initialize the target networks $Q'$ and $\mu'$ with weights $\theta^{Q'}$ $\leftarrow$ $\theta^{Q}$ and $\theta^{\mu'}$ $\leftarrow$ $\theta^{\mu}$;
\State Initialize the auxiliary task network parameters;
\State Initialize replay buffer R;
\For{episode = 1, M}
\State Initialize the Ornstein-Uhlenbeck process for action exploration;
\State Set the initial quantum state as $s_1$;
\For{$t$ = 1, $N_{max}$; $done = false$}
\State Select action $a_t = \mu(\cdot|\theta^\mu)+OU_t$ according to the current policy and exploration noise;
\State Obtain the reward prediction $r_{pre\_t}$ according to the auxiliary task network;
\State Execute action $a_t$ and observe reward $r_t$ and new state $s_{t+1}$  and $done$ signal;
\State Store transition ($s_t$, $a_t$, $r_t$, $s_{t+1}$, $done$, $r_{pre\_t}$) in R;
\State Sample a random minibatch of $N$ transitions ($s_t$, $a_t$, $r_t$, $s_{t+1}$, $done$, $r_{pre\_t}$) from R;
\State Set $y_i = r_i + \gamma Q'(s_{i+1},\mu'(s_{i+1}|\theta^{\mu'})|\theta^{Q'})$;
\State Update the Critic network by minimizing the loss $L = \frac{1}{N} \sum_i (y_i - Q(s_i,a_i|\theta^Q))^2$;
\State Update the Actor network by using the sampled policy gradient
$$
\nabla_{\theta^\mu} J \approx \frac{1}{N} \sum_i \nabla_a Q\left(s, a \mid \theta^Q\right)|_{s=s_i, a=\mu\left(s_i\right)} \nabla_{\theta^\mu} \mu\left(s \mid \theta^\mu\right)|_{s_i};
$$
\State Update the Reward Prediction network by using batch gradient descent $J(\theta_0,\theta_1)=\frac{1}{N} \sum_{i=1}^N (r_{pre\_t}-r_t)^2$;
\State Update the target networks $\theta^{Q'} \leftarrow \tau \theta^Q + (1-\tau)\theta^{Q'}$ and $\theta^{\mu'} \leftarrow \tau \theta^{\mu} + (1-\tau)\theta^{\mu'}$.
\EndFor
\EndFor
\end{algorithmic}
\end{algorithm*}

\section{Numerical Simulation}
In order to verify the performance of the proposed AT-DRL for quantum control, we implement numerical simulations of state preparation on one-qubit, two-qubit, and eight-qubit systems.

\subsection{Parameters Setting}
The neural networks including the Actor network and the Critic network both utilize four hidden layers with 300, 800, 1600, and 800 neurons, respectively. The number of neurons in the output layer is related to the number of control channels in the quantum system. For example, a one-qubit system is controlled by one control channel, so there is only one neuron in the output layer; while eight-qubit systems are controlled by eight channels, so the output layer has eight neurons. Considering that the Critic network aims at guiding
the Actor network, therefore the Critic network must learn faster than the Actor network to ensure the Actor network updates parameters in the correct direction. We set the learning rates of the Actor network and the Critic network as $\alpha_{Actor}=10^{-4}$ and $\alpha_{Critic}=10^{-3}$, respectively. In the simulation, the discount factor is taken as $\gamma = 0.99$ and the parameters are optimized by Adam's algorithm with the batch size $B = 64$. The action space is defined in $[-1,1]$, which means that the action can output any real number between $-1$ and $1$. In the whole network setting, we use a six-layer neural network with an input layer, an output layer, and four hidden layers. In the comparison with the DQN method, we use an input layer, an output layer and one hidden layer, with the hidden layer containing 10 neurons. The learning rate is set as $\alpha_{DQN}=0.01$, and the batch size is set to be $B = 32$ \cite{53}. The action space is the same as the AT-DRL method.

Before inputting quantum states (that are complex vectors) into the neural networks, we first transform them into real vectors using the following equation \cite{31}:
\begin{equation}\nonumber
\begin{gathered}
s_j=\left[\Re\left(\left\langle 0 \mid \psi_j\right\rangle\right), \Re\left(\left\langle 1 \mid \psi_j\right\rangle\right), \ldots, \Re\left(\left\langle n-1 \mid \psi_j\right\rangle\right),\right. \\
\left.\quad \Im\left(\left\langle 0 \mid \psi_j\right\rangle\right), \Im\left(\left\langle 1 \mid \psi_j\right\rangle\right), \ldots, \Im\left(\left\langle n-1 \mid \psi_j\right\rangle\right)\right],
\end{gathered}
\end{equation}
where $\{|k\rangle\}_{k=0}^{n-1}$ is a computational basis of the $n$-qubit system, $\Re(\cdot)$ and $\Im(\cdot)$ denote the real and imaginary parts of a complex number.

\subsection{One-qubit Quantum Systems}
We consider the one-qubit quantum system \cite{53}, whose Hamiltonian is
\begin{equation}
\begin{aligned}
 H[J(t)]=4 J(t) \sigma_z+h \sigma_x,
 \end{aligned}
\end{equation}
where $\sigma_x$ and $\sigma_z$ are Pauli matrices, i.e.,
$$
\sigma_x=\left(\begin{array}{ll}
0 & 1 \\
1 & 0
\end{array}\right), \quad \sigma_y=\left(\begin{array}{cc}
0 & -i \\
i & 0
\end{array}\right), \quad \sigma_z=\left(\begin{array}{cc}
1 & 0 \\
0 & -1
\end{array}\right).
$$
The whole Hamiltonian describes a singlet-triplet qubit or a single spin with energy gap $h$ under tunable control fields \cite{53}. To facilitate calculation, $h = 1$ is specified. The goal is to find an appropriate control $J(t)$ so that the system evolves from the initial state $\ket{\psi_0}$ to the target state $\ket{\psi_f}$ with a high fidelity. In the simulation, assume fixed evolution time for each step is $\pi / 20$.

For the one-qubit system, good performance can be attained by only using the DDPG algorithm. Thus, we only use guided rewards and do not utilize the auxiliary task. In particular, we consider target states with two cases: an Eigenstate as the target state and a Superposition state as the target state.

\subsubsection{Eigenstates Preparation}
\begin{figure*}[!t]
\centering
\subfloat[]{\includegraphics[width=2.6in]{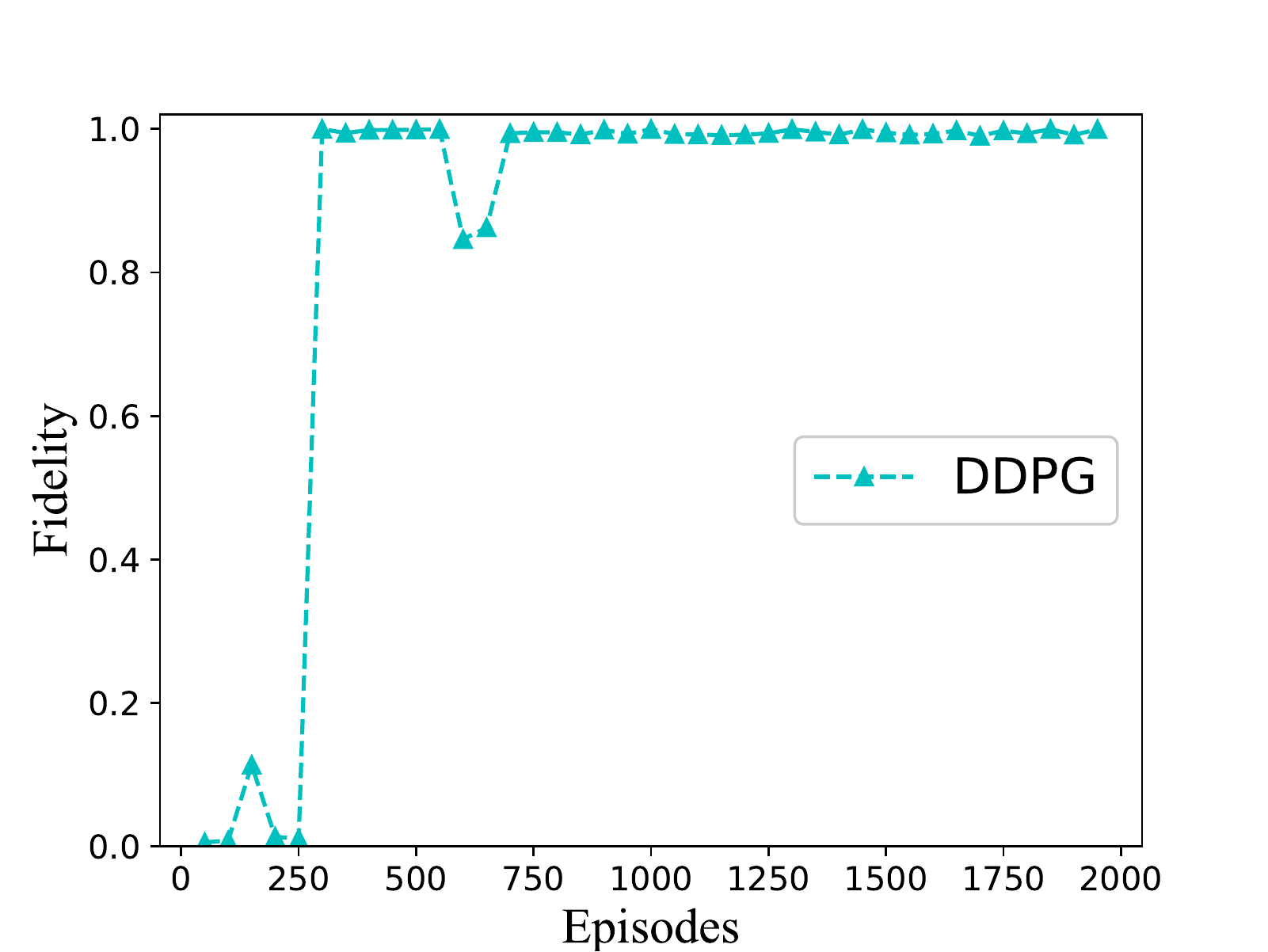}%
\label{fig_first_case}}
\hfil
\subfloat[]{\includegraphics[width=2.6in]{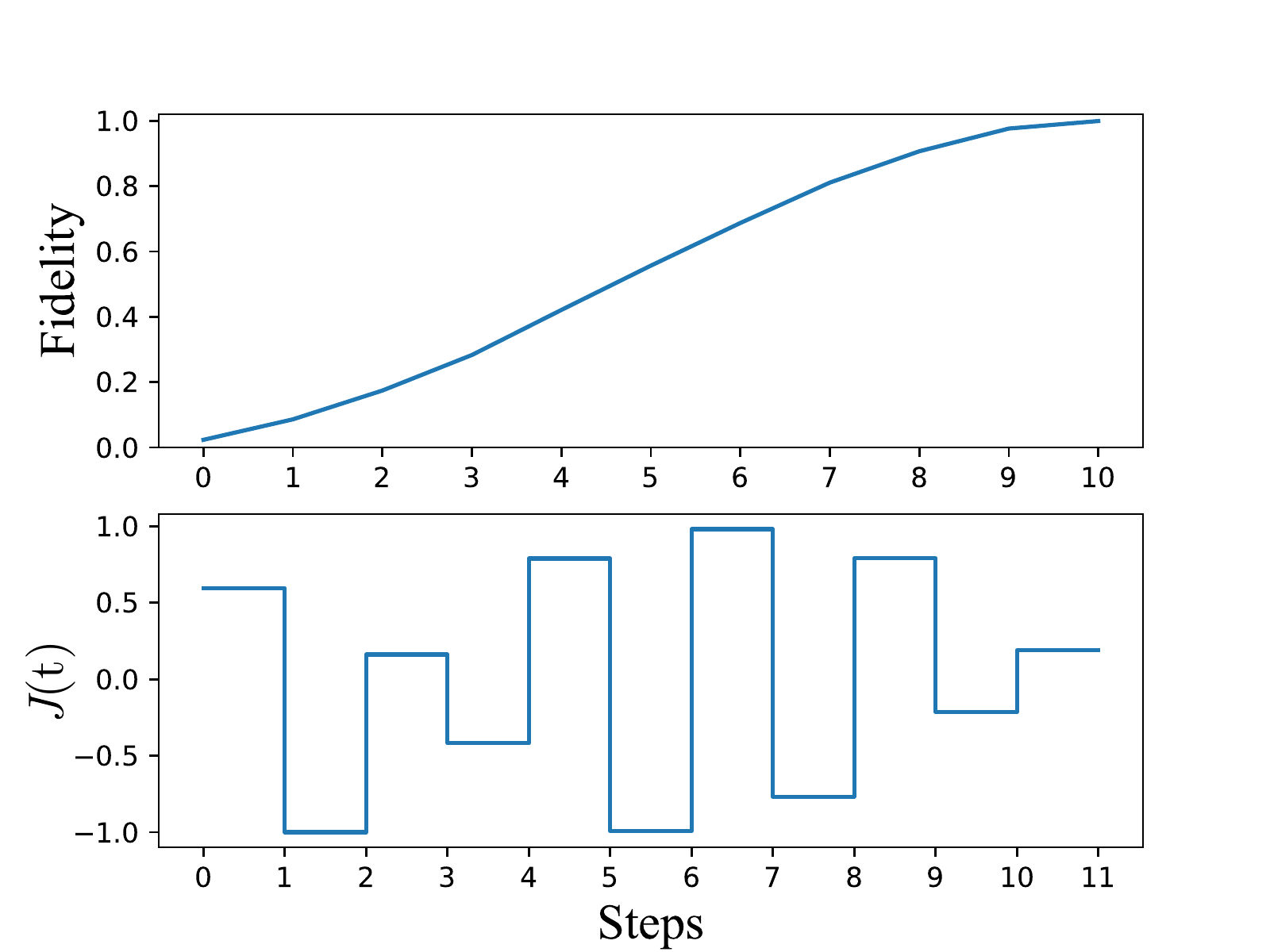}%
\label{fig_second_case}}
\hfil
\subfloat[]{\includegraphics[width=1.9in]{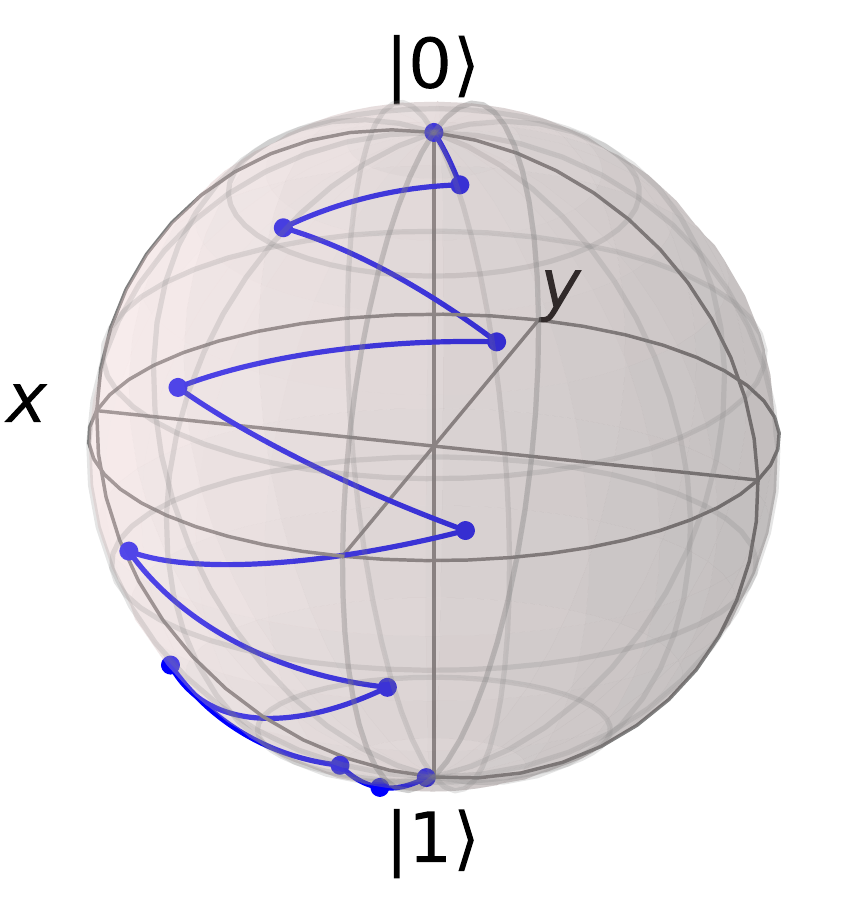}%
\label{fig_second_case}}
\caption{Preparation of $\ket{0}$ on a one-qubit quantum system. (a) the whole learning process, (b) the fidelity and control laws of each step within one episode, and (c) the evolution trajectory on the Bloch sphere.}
\label{fig_sim}
\end{figure*}

\begin{figure*}[!t]
\centering
\subfloat[]{\includegraphics[width=2.6in]{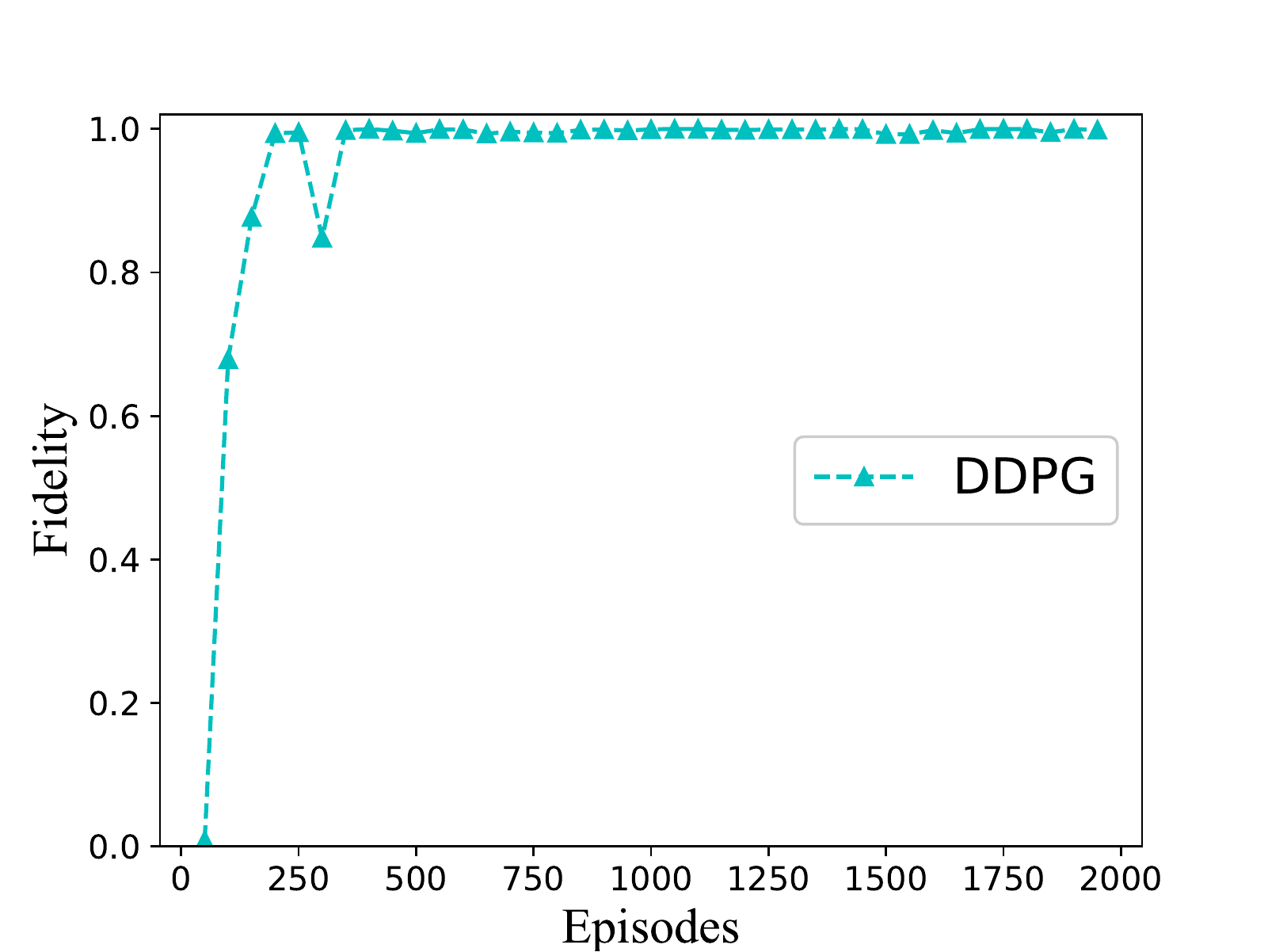}%
\label{fig_first_case}}
\hfil
\subfloat[]{\includegraphics[width=2.6in]{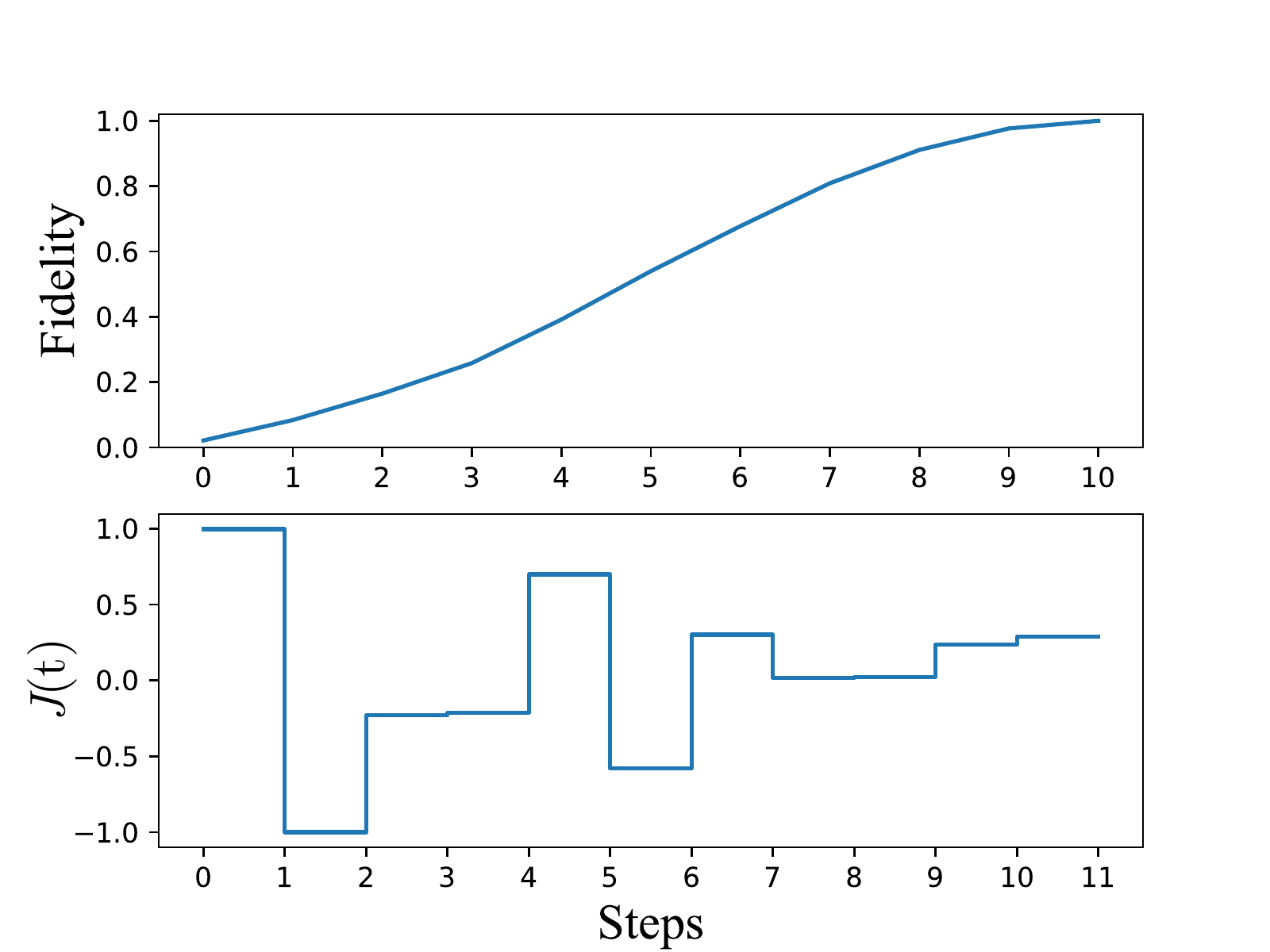}%
\label{fig_second_case}}
\hfil
\subfloat[]{\includegraphics[width=1.9in]{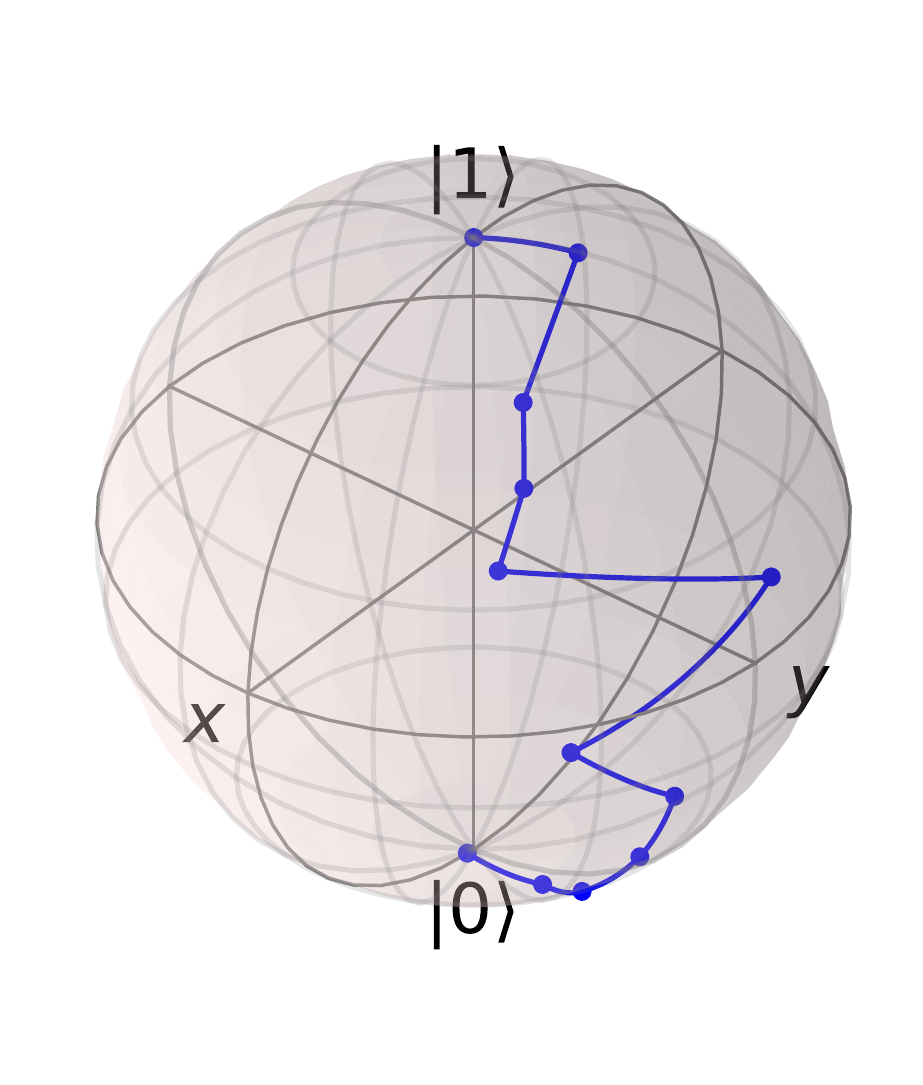}%
\label{fig_second_case}}
\caption{Preparation of $\ket{1}$ on a one-qubit quantum system. (a) the whole learning process, (b) the fidelity and control laws of each step within one episode, and (c) the evolution trajectory on the Bloch sphere.}
\label{fig_sim}
\end{figure*}

As shown in Fig. 2, the initial state is $\ket{1}$ and the target state is $\ket{0}$. Considering the results in Fig. 2(a), the fidelity of the system reaches a high value at the 300--th episode and is around 0.9993 at the end (taking the average of the last 100 episodes). According to Fig. 3(a), the initial state is $\ket{0}$ and the target state is $\ket{1}$. the system is manipulated to a state with a fidelity of 0.9976 against the target state $\ket{1}$. Figs. 2(b) and 3(b) show the fidelity and the action values obtained from  the Action network, demonstrating that our method finds a control strategy with 11 control pulses and achieves a fidelity above 0.99. Fig. 2(c) and Fig. 3(c) visualize the trajectories of the state in Bloch representations \cite{lou2011state}.

\subsubsection{Superposition States preparation}
\begin{figure*}[!t]
\centering
\subfloat[]{\includegraphics[width=2.6in]{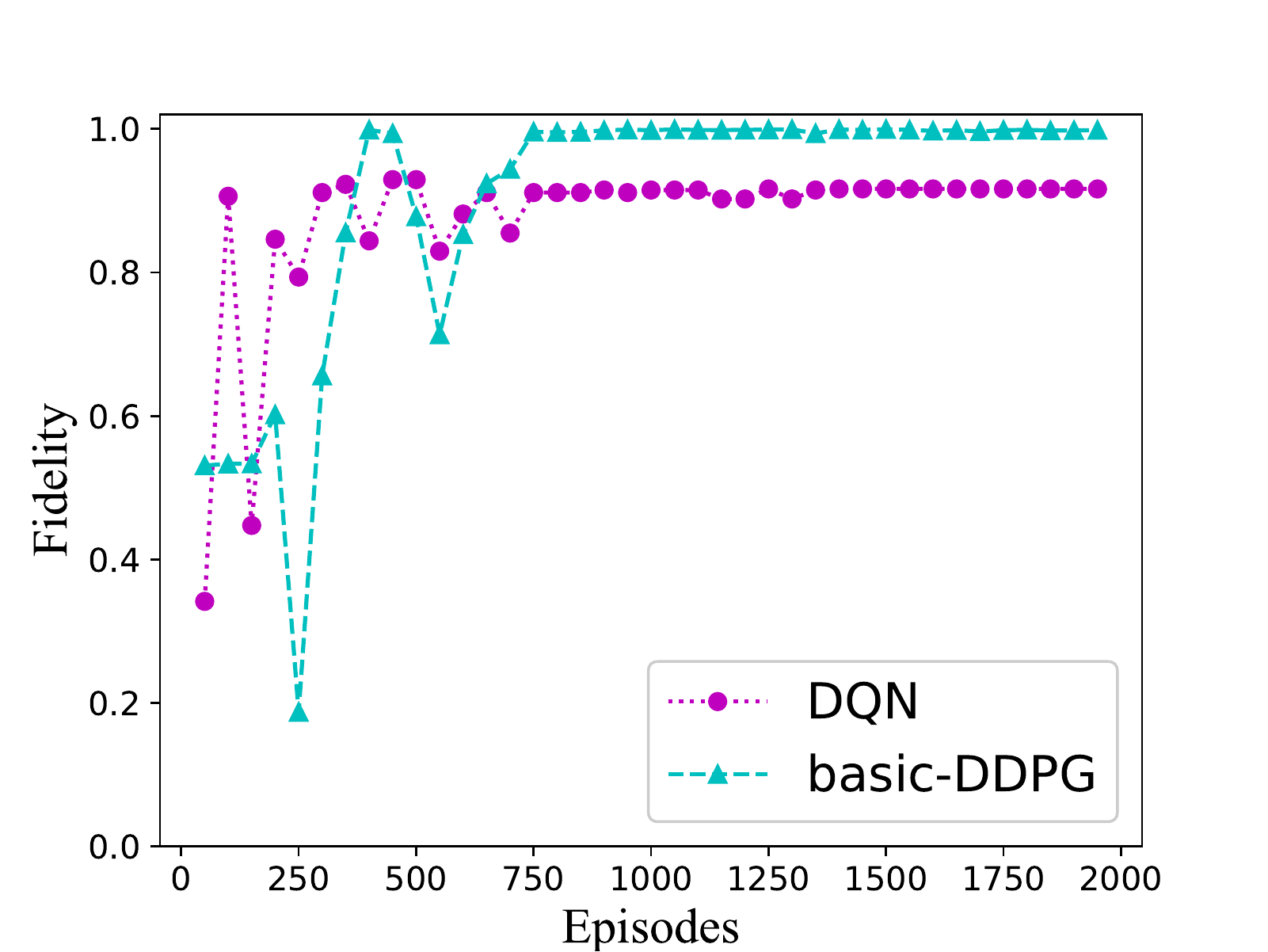}%
\label{fig_first_case}}
\hfil
\subfloat[]{\includegraphics[width=2.6in]{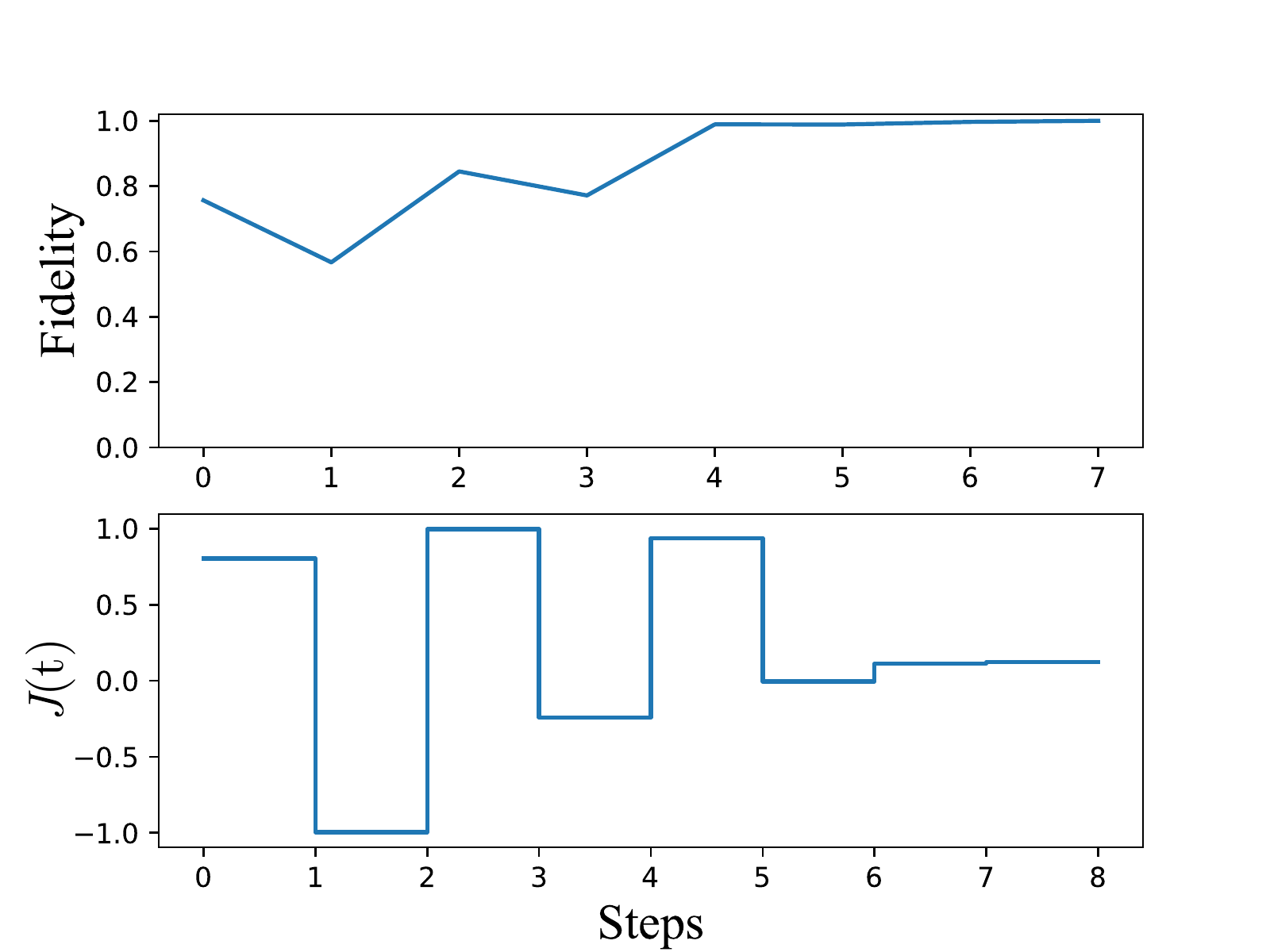}%
\label{fig_second_case}}
\hfil
\subfloat[]{\includegraphics[width=1.9in]{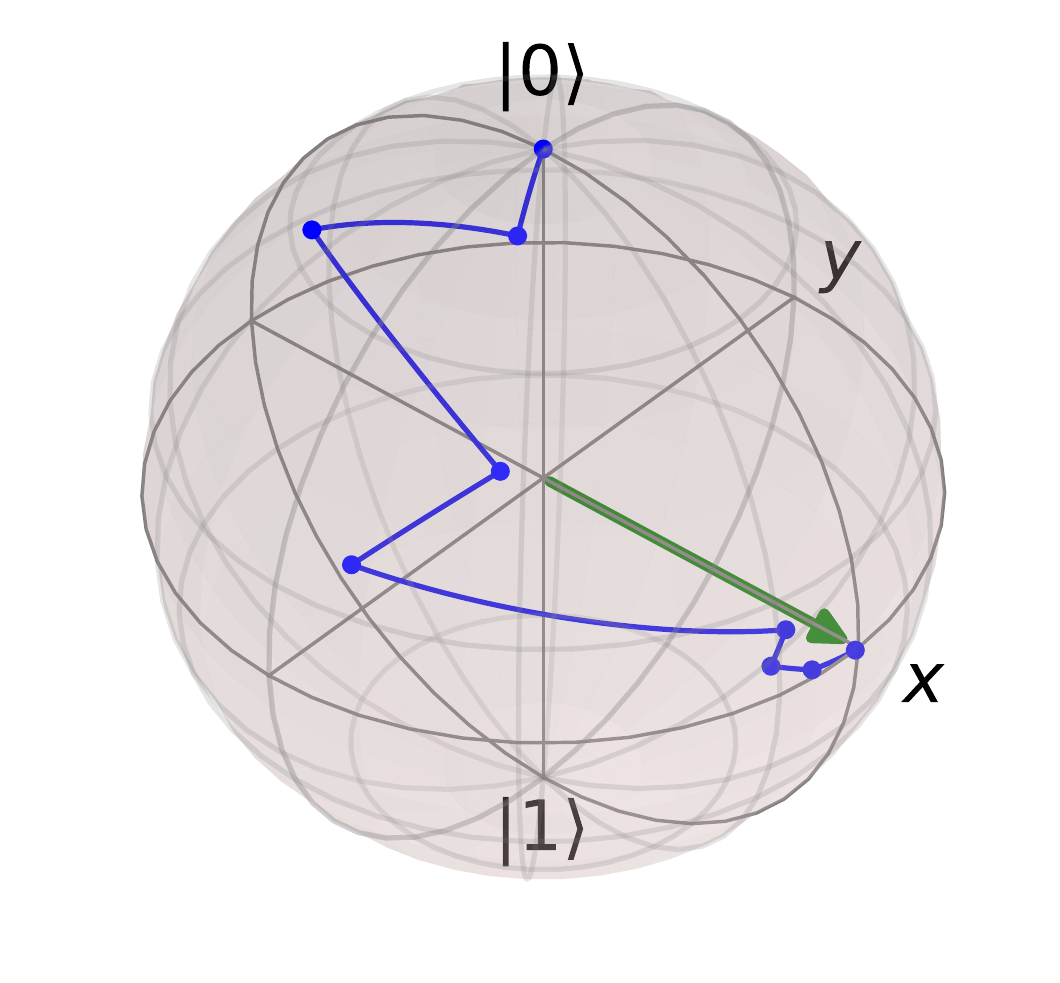}%
\label{fig_second_case}}
\caption{Preparation of superposition state in a one-qubit quantum system with the initial state $\ket{1}$. (a) the whole learning process, (b) the fidelity and control laws of each step within one episode, and (c) the evolution trajectory on the Bloch sphere.}
\label{fig_sim}
\end{figure*}

\begin{figure*}[!t]
\centering
\subfloat[]{\includegraphics[width=2.6in]{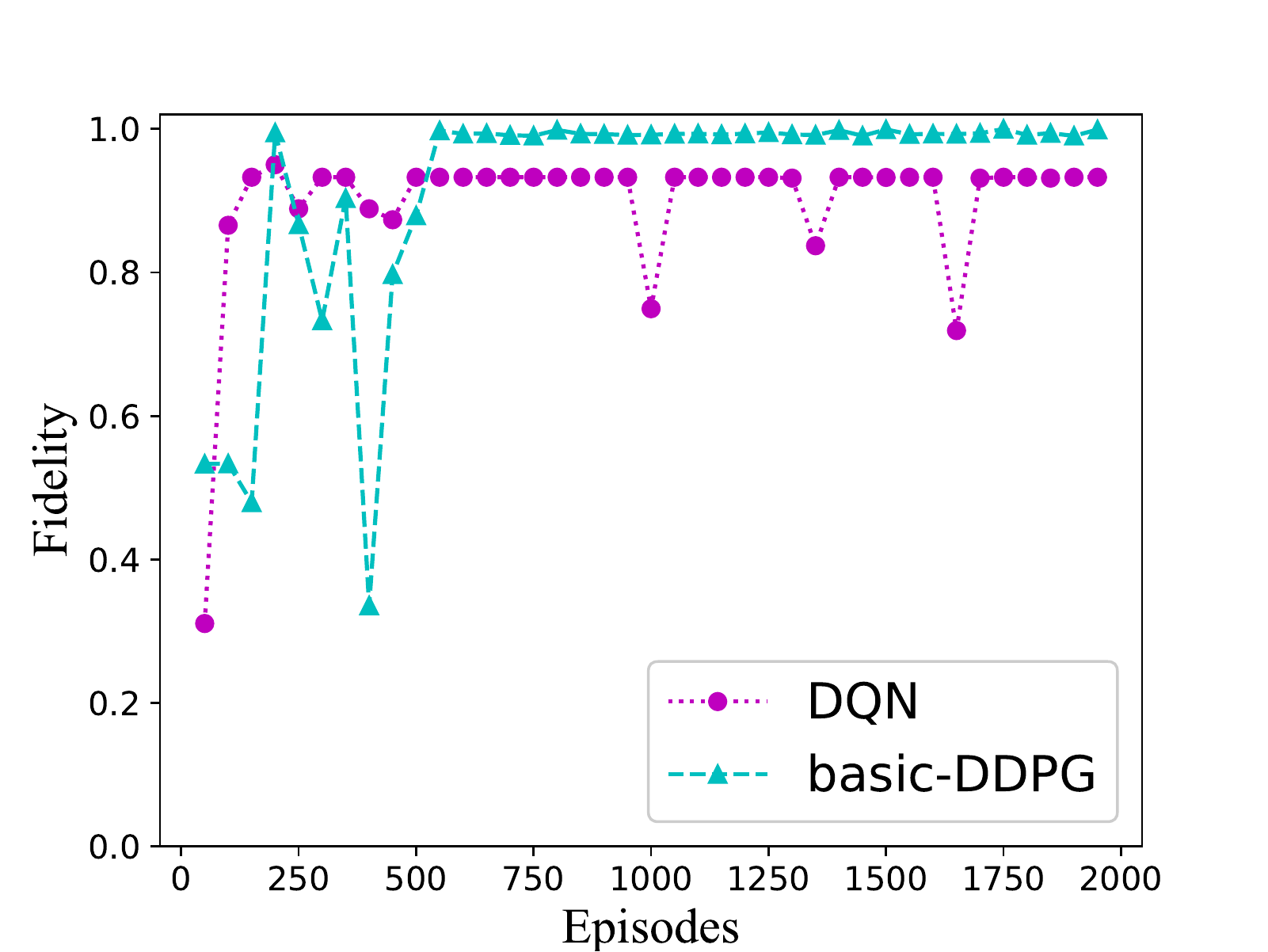}%
\label{fig_first_case}}
\hfil
\subfloat[]{\includegraphics[width=2.6in]{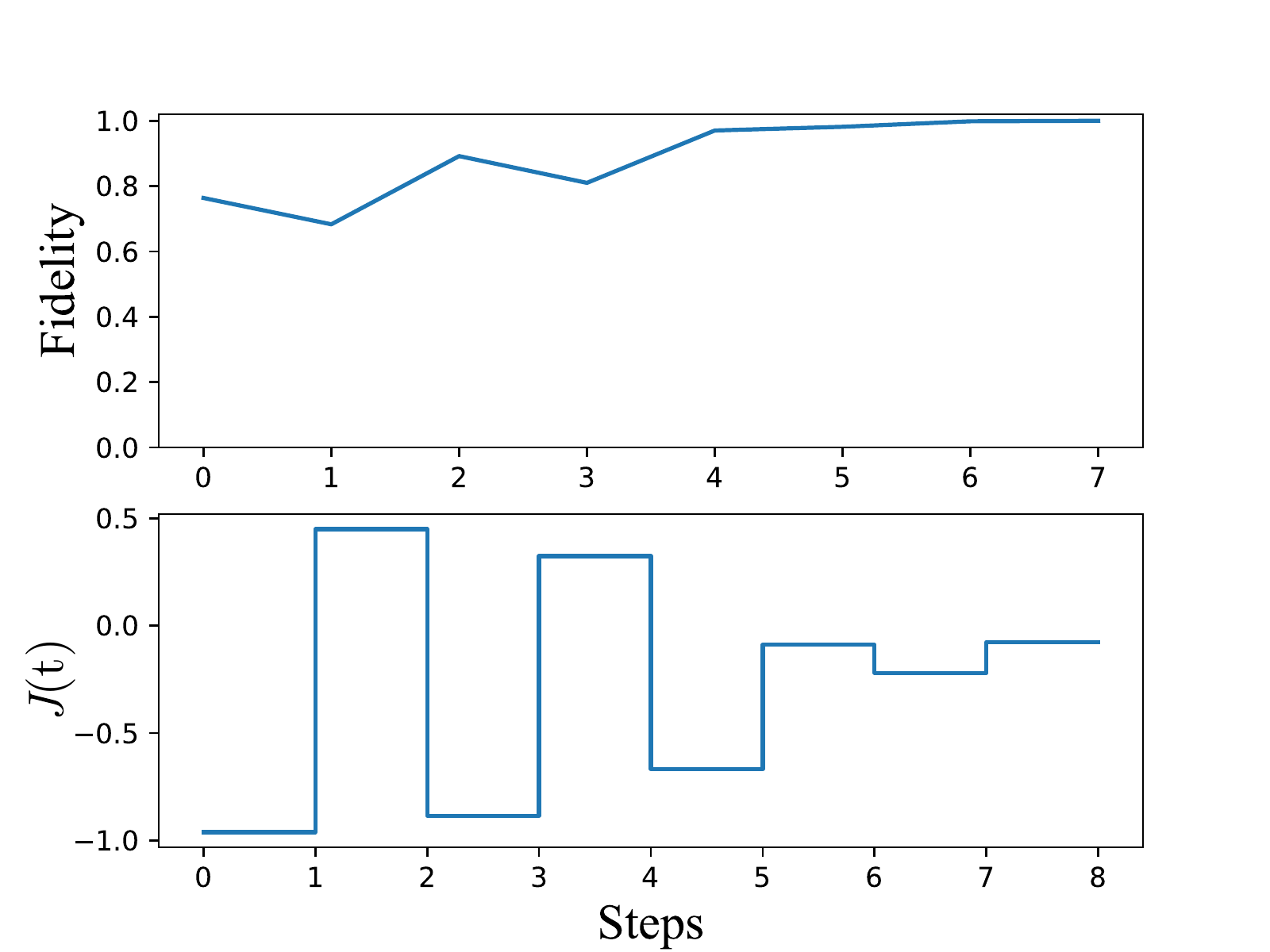}%
\label{fig_second_case}}
\hfil
\subfloat[]{\includegraphics[width=1.9in]{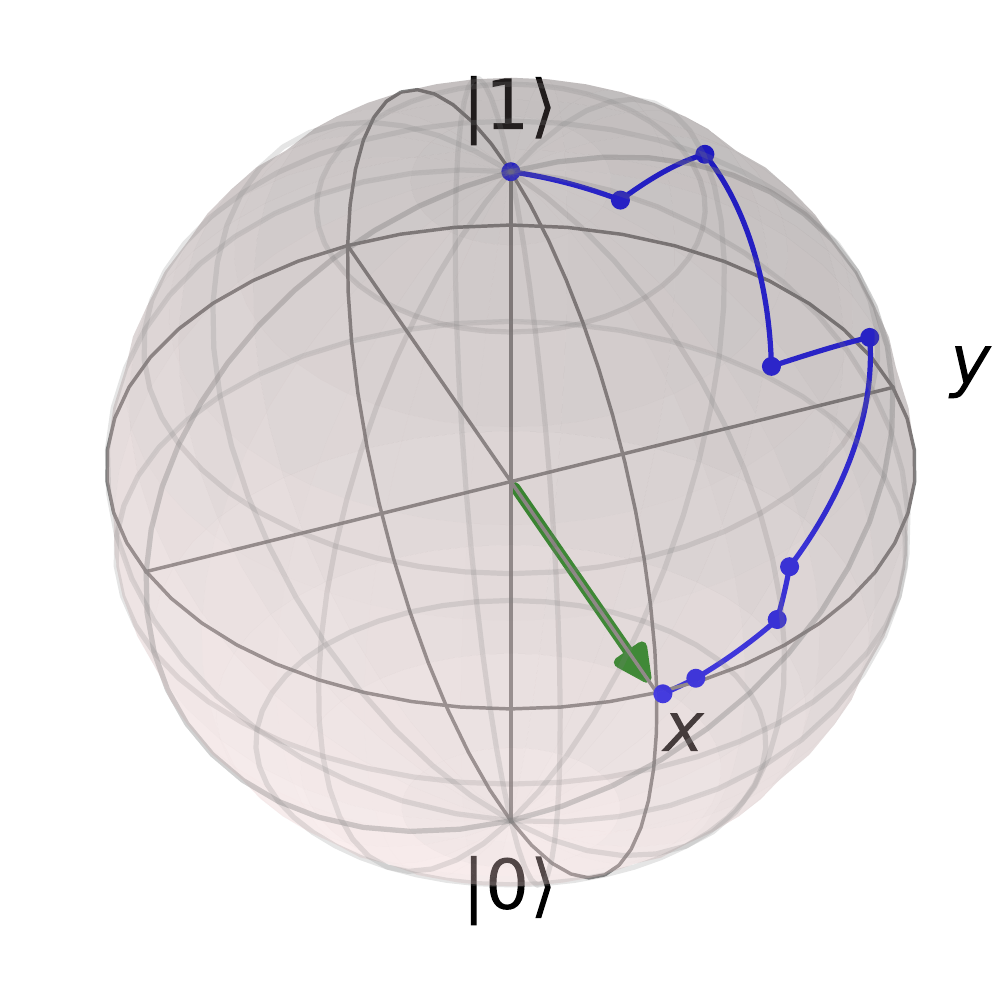}%
\label{fig_second_case}}
\caption{Preparation of superposition state in a one-qubit quantum system with the initial state $\ket{0}$. (a) the whole learning process, (b) the fidelity and control laws of each step within one episode, and (c) the evolution trajectory on the Bloch sphere.}
\label{fig_sim}
\end{figure*}

We also implement two simulations, the first one with the initial state $\ket{1}$ and the target state $(\frac{1}{2}+\frac{1}{2} i)\ket{1} + (\frac{1}{2}+\frac{1}{2} i)\ket{0}$, and the second one with the initial state $\ket{0}$ and the target state $(\frac{1}{2}+\frac{1}{2} i)\ket{1} + (\frac{1}{2}+\frac{1}{2} i)\ket{0}$. The simulation results are shown in Fig. 4 and Fig. 5. From Fig. 4(a), both the DDPG algorithm and the DQN algorithm gradually converge for around 750 episodes with different final fidelities. The DDPG algorithm can reach about 0.9991, while the DQN algorithm can only achieve about 0.9163, demonstrating that the DDPG algorithm outperforms the DQN algorithm. Similar simulation results occur in Fig. 5(a), where the fidelity can reach 0.9941 through the DDPG algorithm, while the DQN algorithm can only reach 0.9327. In addition, the control system obtained by the DQN algorithm has unstable oscillations compared with the control obtained by the DDPG algorithm.

From the above simulations, the DDPG algorithm performs a more accurate control effect than the DQN algorithm for one-qubit state preparation.

\subsection{Two-qubit Quantum Systems}
Now, we consider a two-qubit quantum system, with its Hamiltonian as \cite{53}
$$H(t)=H_0 + u_1(t) H_1+u_2(t)H_2,$$
where the free Hamiltonian is
$H_0=\sigma_x \otimes \sigma_x + \sigma_y \otimes \sigma_y $, and the control Hamiltonians are $H_1=\sigma_z^{(1)} \otimes I_2$ and $H_2= I_2\otimes \sigma_z^{(2)}$, with $I_2$ denotes the identity operator for two-level systems. The control laws  $u_1(t)$ and $u_2(t)$ take values in [-1,1]. The evolution step size of the two-qubit system is set to be $\pi / 40$. In the simulation, we compare the AT-DRL algorithm with the basic DDPG and DQN algorithms.

\subsubsection{Eigenstate Preparation}
We choose two target states $s_1 = \ket{00}$ and $s_2 = \ket{11}$ to verify the effectiveness of the algorithm.
\begin{figure*}[!t]
\centering
\subfloat[]{\includegraphics[width=3.5in]{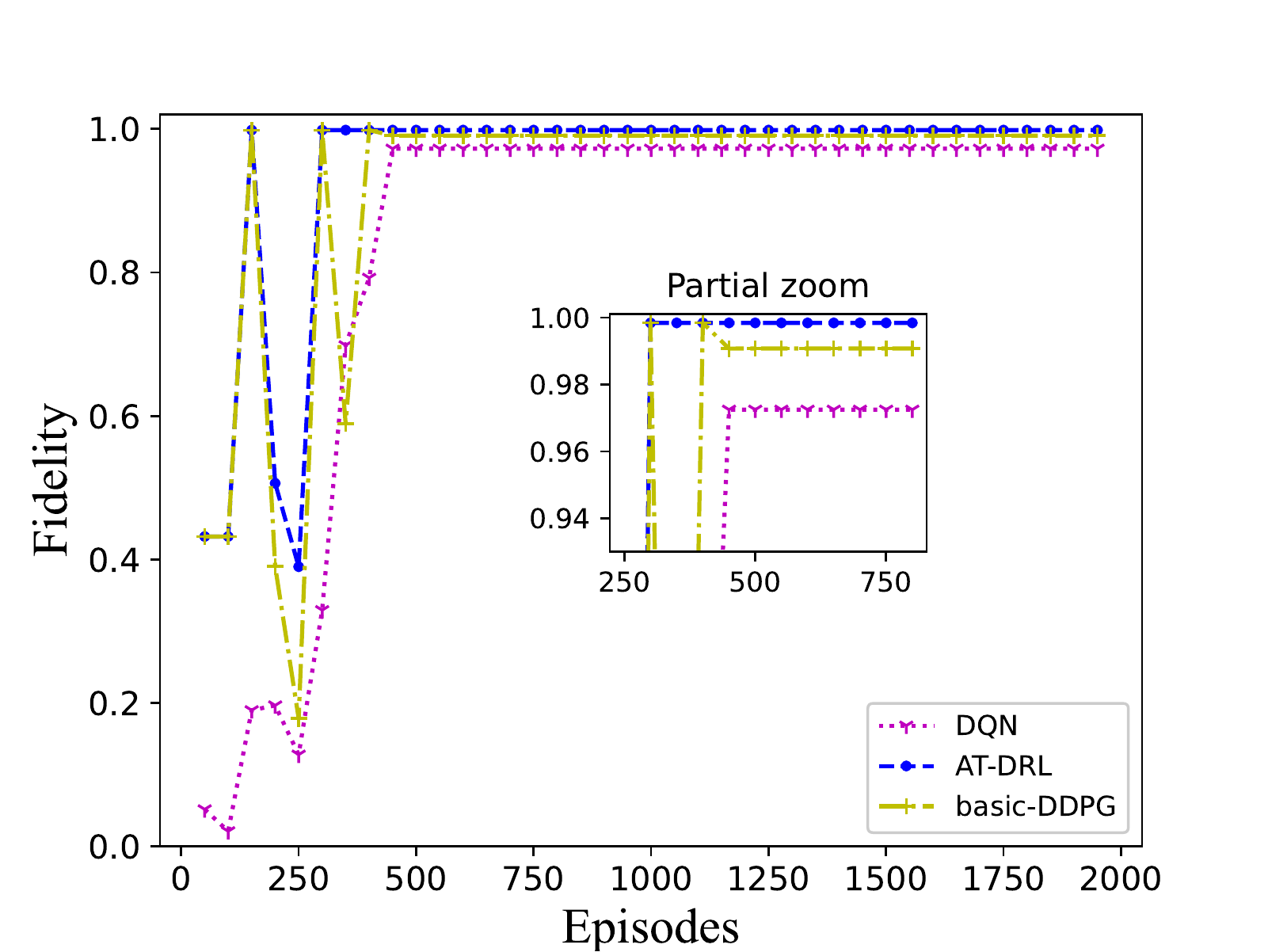}%
\label{fig_first_case}}
\hfil
\subfloat[]{\includegraphics[width=3.5in]{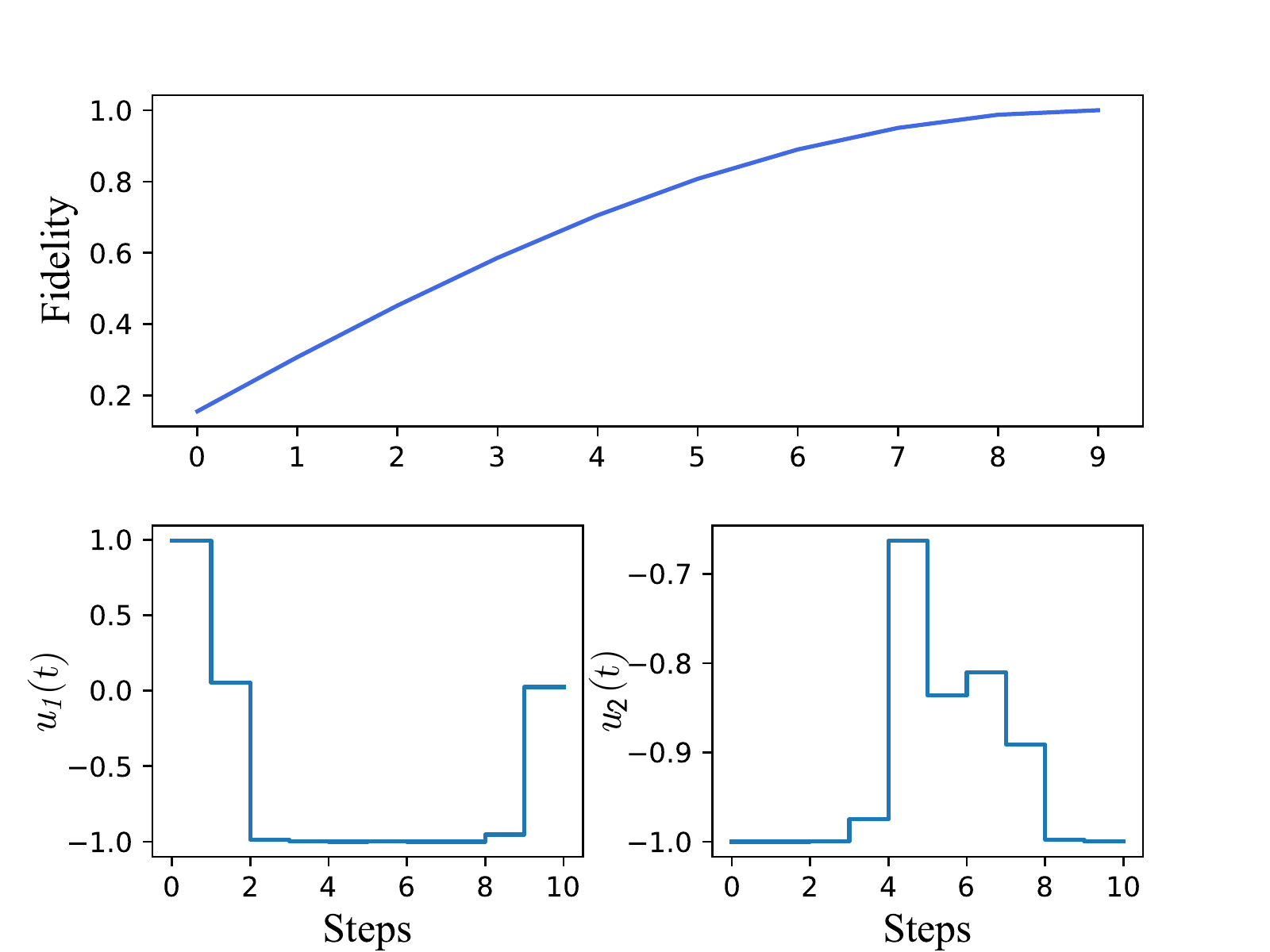}%
\label{fig_second_case}}
\caption{Preparation of eigenstate in a two-qubit quantum system with the initial state  $\ket{00}$. (a) the whole learning process, (b) the fidelity and control laws of each step within one episode.}
\label{fig_sim}
\end{figure*}

\begin{figure*}[!t]
\centering
\subfloat[]{\includegraphics[width=3.5in]{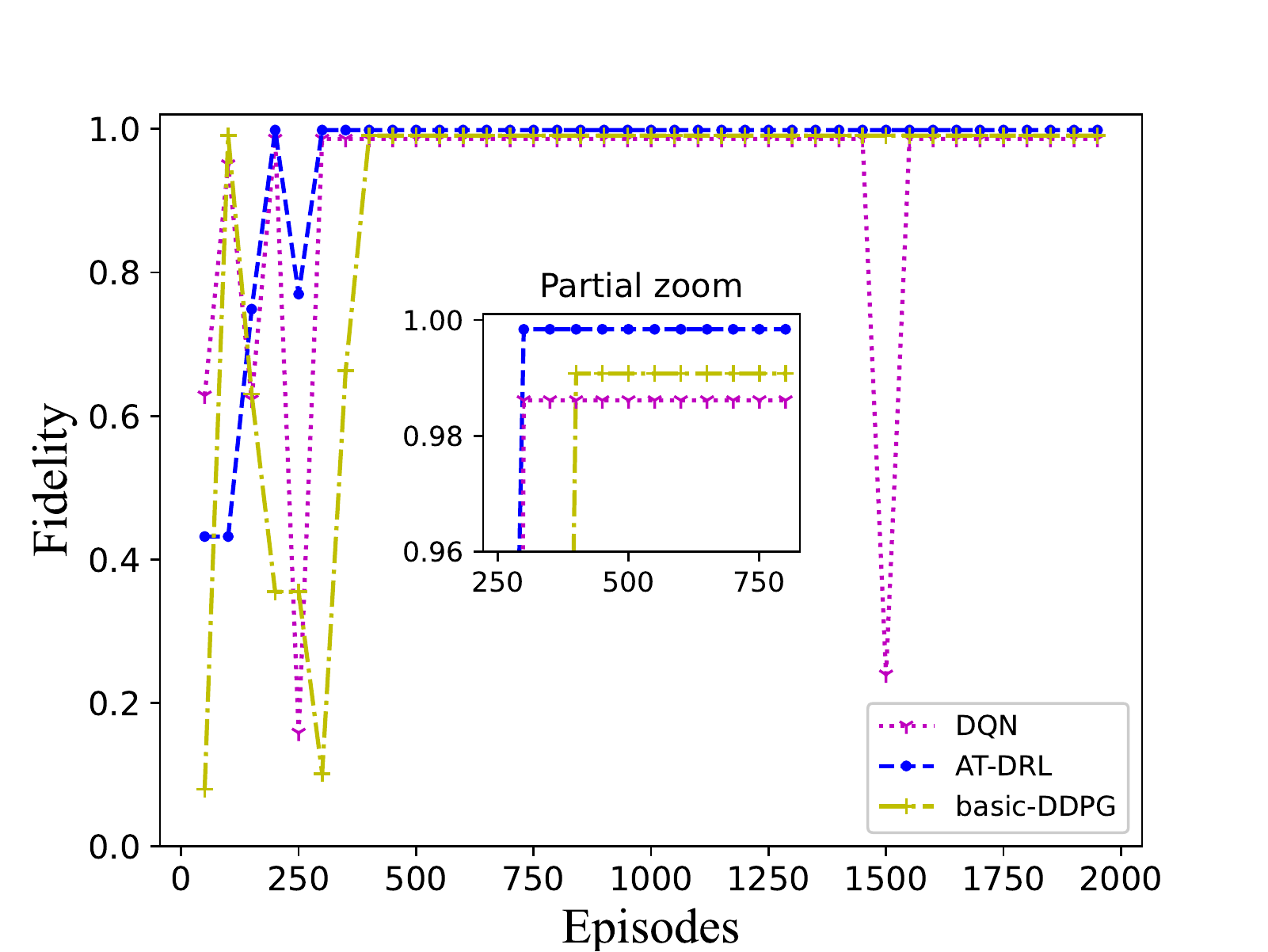}%
\label{fig_first_case}}
\hfil
\subfloat[]{\includegraphics[width=3.5in]{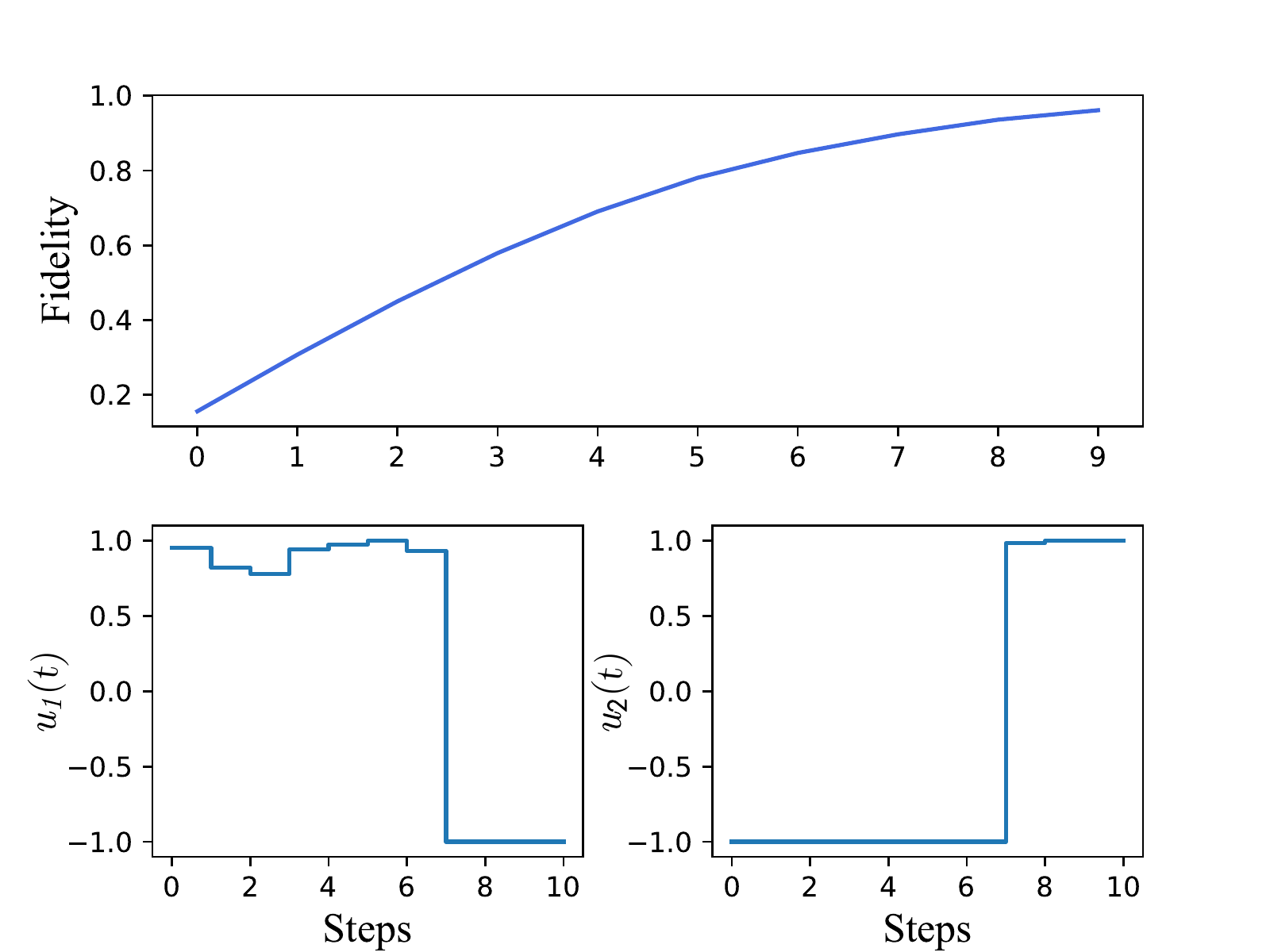}%
\label{fig_first_case}}
\caption{Preparation of eigenstate in a two-qubit quantum system with the initial state $\ket{11}$. (a) the whole learning process, (b) the fidelity and control laws of each step within one episode.}
\label{fig_sim}
\end{figure*}

The simulation results are shown in Fig. 6 and Fig. 7, where Fig. 6 corresponds to the state transfer from $s_1$ to $s_2$, and Fig. 7 corresponds to the state transfer from $s_2$ to $s_1$. Based on Fig. 6(a) and simulation data, AT-DRL achieves a stable fidelity of 0.9984 after the 300-th episode, while the basic DDPG only achieves a final fidelity of 0.9907. DQN is much inferior to the other two algorithms, as it can only achieve a final fidelity of 0.97 with longer training episodes. Similar to Fig. 6(a), AT-DRL in Fig. 7(a) achieves better results than the two other algorithms.

\subsubsection{Bell State Preparation}
The Bell states are the maximally entangled states of two-qubit systems with the maximum entanglement degree and are also the powerful resource for constituting quantum communication \cite{54}. We consider Bell states, which are usually difficult to generate and also important for various applications.  In the simulations, we demonstrate the effectiveness of the algorithm by preparing two states $\ket{\phi^+}$ and $\ket{\phi^-}$ in (4).
\begin{figure*}[!t]
\centering
\subfloat[]{\includegraphics[width=3.5in]{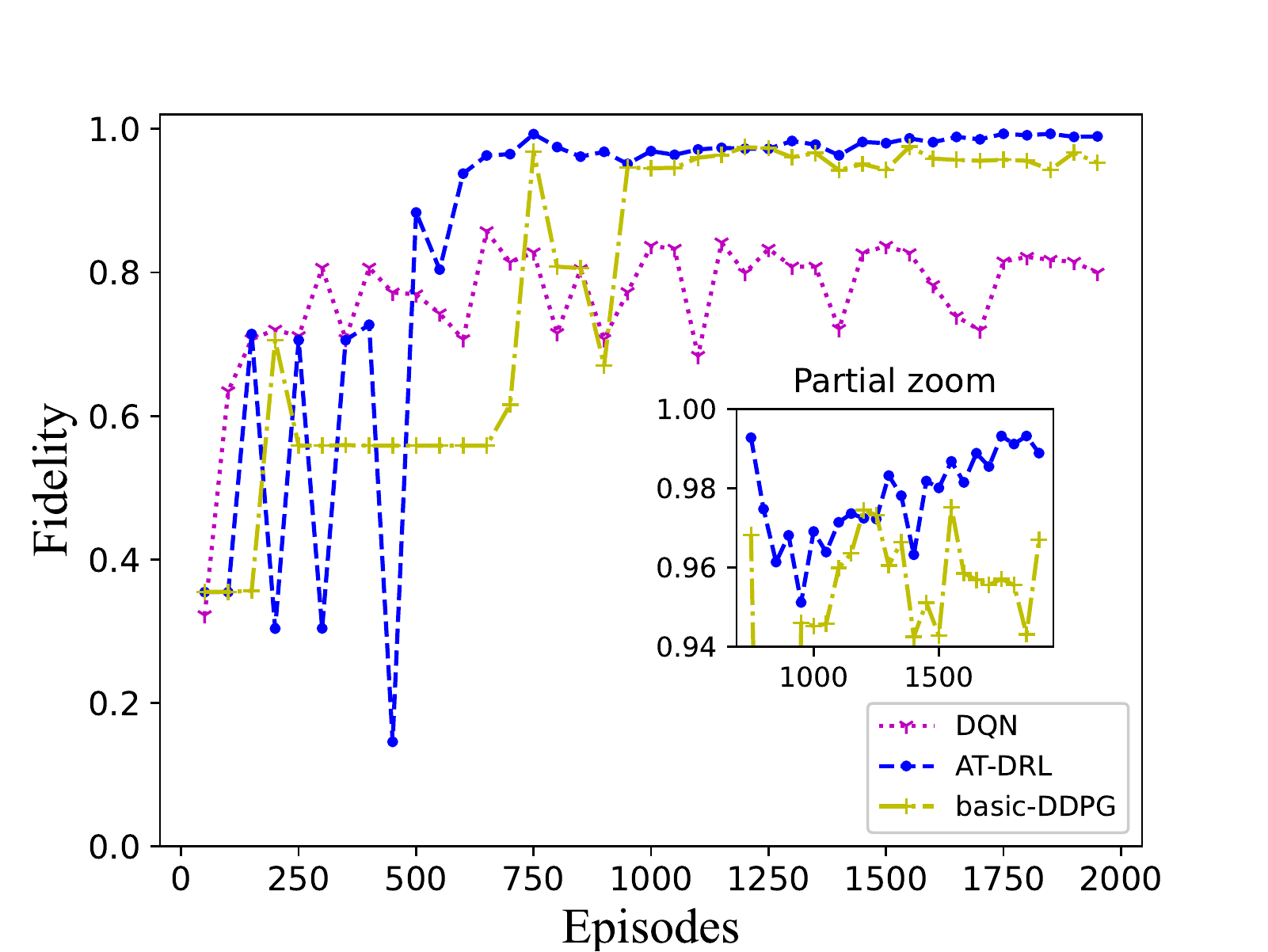}%
\label{fig_first_case}}
\hfil
\subfloat[]{\includegraphics[width=3.5in]{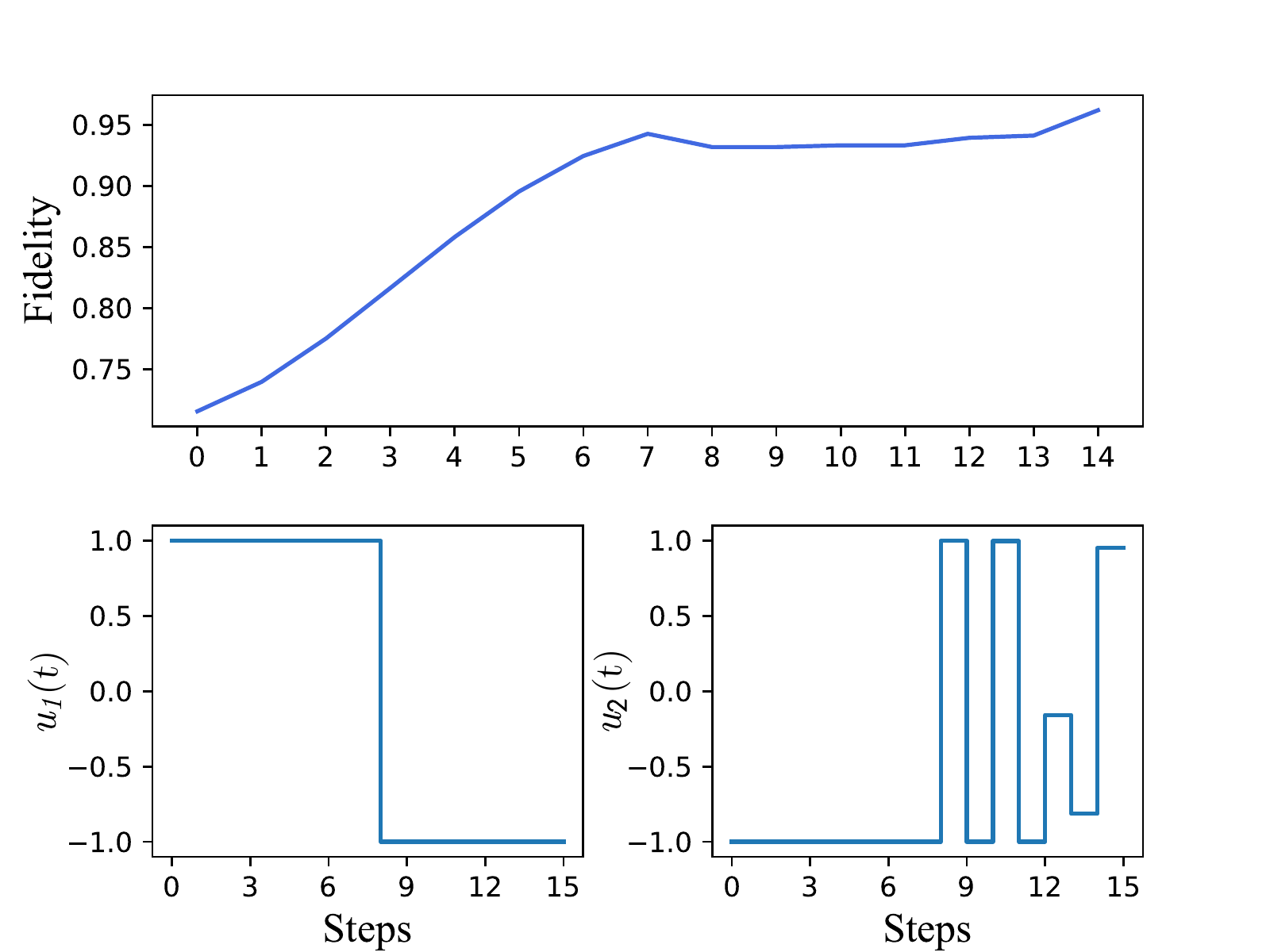}%
\label{fig_second_case}}
\caption{Preparation of $\ket{\phi^+}$ on a two-qubit quantum system. (a) the whole learning process, and (b) the fidelity and control laws of each step within one episode.}
\label{fig_sim}
\end{figure*}

\begin{figure*}[!t]
\centering
\subfloat[]{\includegraphics[width=3.5in]{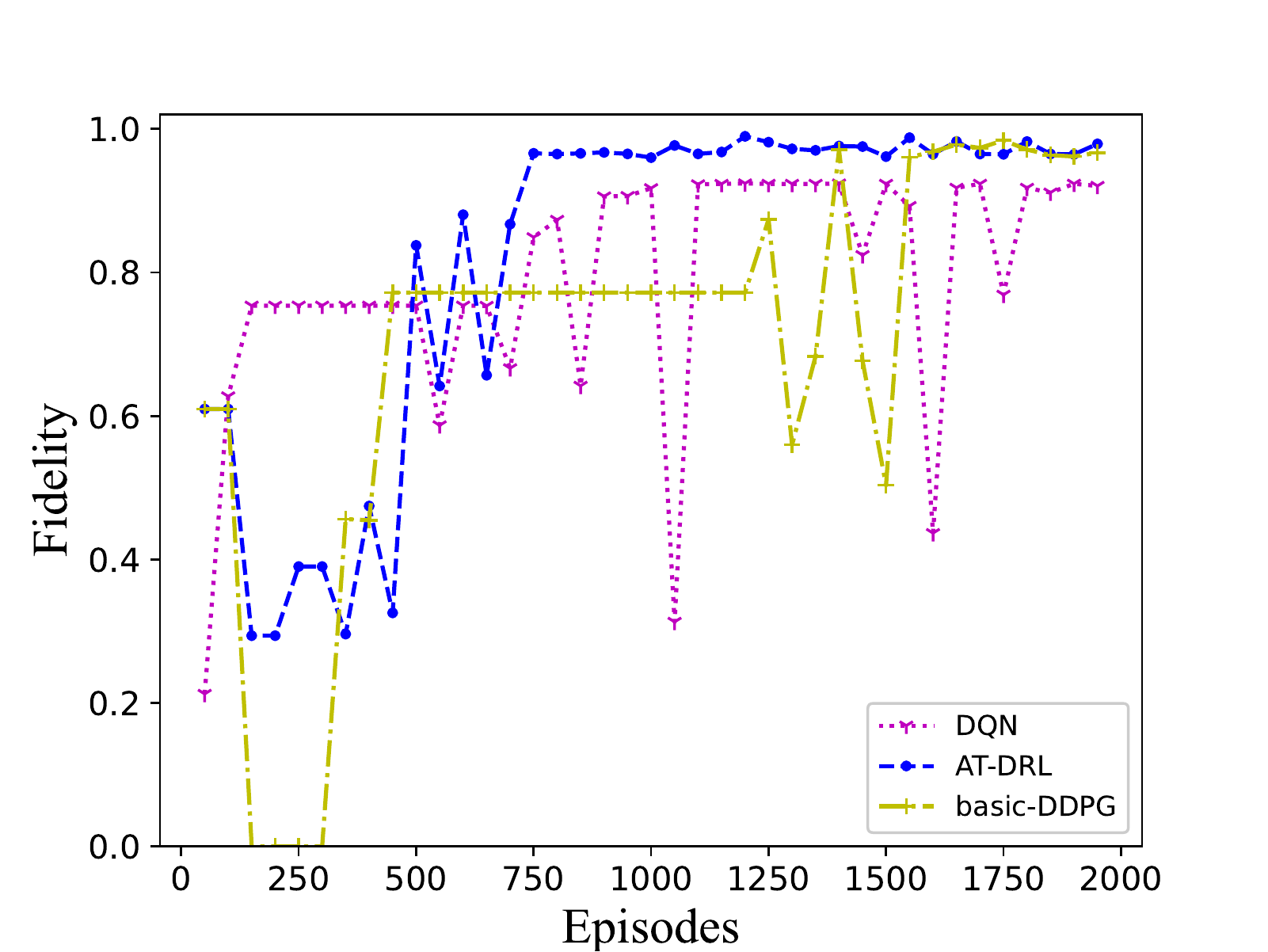}%
\label{fig_first_case}}
\hfil
\subfloat[]{\includegraphics[width=3.5in]{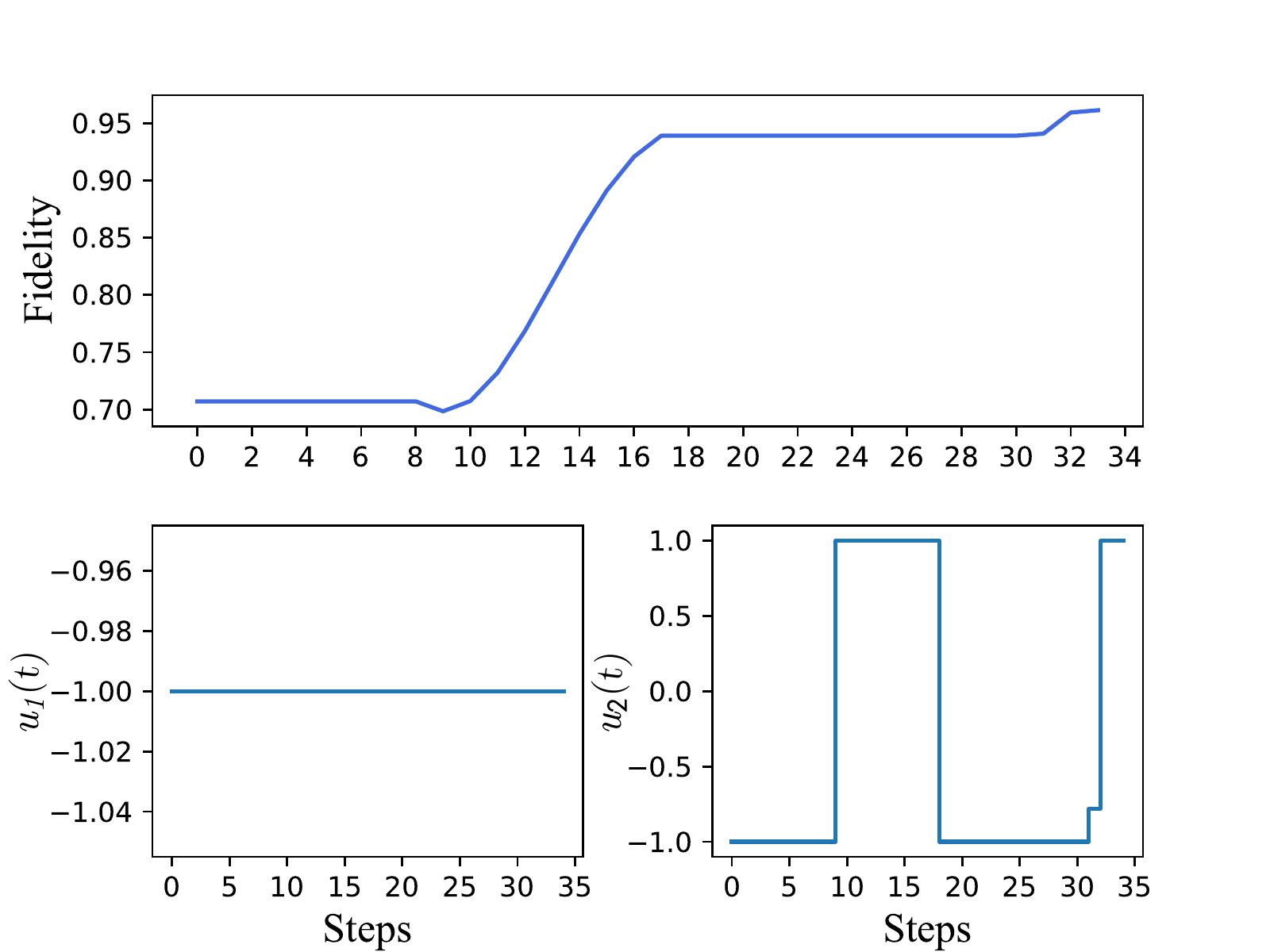}%
\label{fig_first_case}}
\caption{Preparation of $\ket{\phi^-}$ on a two-qubit quantum system. (a) the whole learning process, and (b) the fidelity and control laws of each step within one episode.}
\label{fig_sim}
\end{figure*}

The simulation results are shown in Fig. 8 and Fig. 9, where Fig. 8 represents the preparation of the Bell state $\ket{\phi^+}=\frac{1}{\sqrt{2}}\ket{00}+\frac{1}{\sqrt{2}}\ket{11}$ from the state $s_1$. As we can see, the blue and yellow lines stabilize above 0.9, indicating that both the basic DDPG algorithm and the AT-DRL algorithm have good performance regarding the average of the last five results, where the basic DDPG algorithm reaches 0.955 and the AT-DRL algorithm reaches 0.991. However, the DQN algorithm can only reach 0.813. In terms of convergence rate, the AT-DRL algorithm achieves a fidelity of greater than 0.90 at the 600-th episode, which is a significant improvement. Fig. 9 represents the preparation of the Bell state $\ket{\phi^-}=\frac{1}{\sqrt{2}}\ket{01}+\frac{1}{\sqrt{2}}\ket{10}$ from $s_1$. The advantage of the AT-DRL algorithm can be intuitively demonstrated by its rapid convergence to 0.971. In comparison, the basic DDPG algorithm takes more episodes to realize a stable learning curve in Fig. 9(a). The DQN algorithm can only reach a fidelity of 0.92 and has large fluctuations in the learning process.

\subsection{Multi-qubit Systems}
We conduct simulations on a spin-transfer system \cite{53}, whose Hamiltonian is
\begin{equation}
\begin{aligned}
H(t)= & C \sum_{k=1}^{K-1} (I_2^{\otimes (k-1)} \otimes S_x^{(k)} \otimes S_x^{(k+1)} \otimes I_2^{\otimes (K-k-1)} \\  &+I_2^{\otimes (k-1)} \otimes S_y^{(k)} \otimes S_y^{(k+1)} \otimes I_2^{\otimes (K-k-1)}) \\
& +\sum_{k=1}^K 2 B_k(t) I_2^{\otimes (k-1)} \otimes S_z^{(k)} \otimes I_2^{\otimes (K-k)},
\end{aligned}
\end{equation}
where $K$ is the total number of spins, $s_x^{(k)}$, $s_y^{(k)}$ and $s_z^{(k)}$ are the $k$-th spin operators. $C$ describes the constant nearest-neighbor coupling strength and is set as $C=1$ in this work. $B_k(t)$ is the time-dependent local magnetic field applied on the $k$th spin. We consider the spin transfer of an eight-qubit system, i.e., $K=8$.  As is demonstrated in Fig. 10, the goal is to evolve the system from an initial state $\ket{10000000}$ to the target state $\ket{00000001}$. The status $\ket{10000000}$ represents that the leftmost qubit is spin-up and the remaining seven qubits are spin-down, while the status $\ket{00000001}$ represents that the rightmost qubit is spin-up and the remaining seven qubits is spin-down.
\begin{figure}[!t]
\centering
{\includegraphics[width=3in]{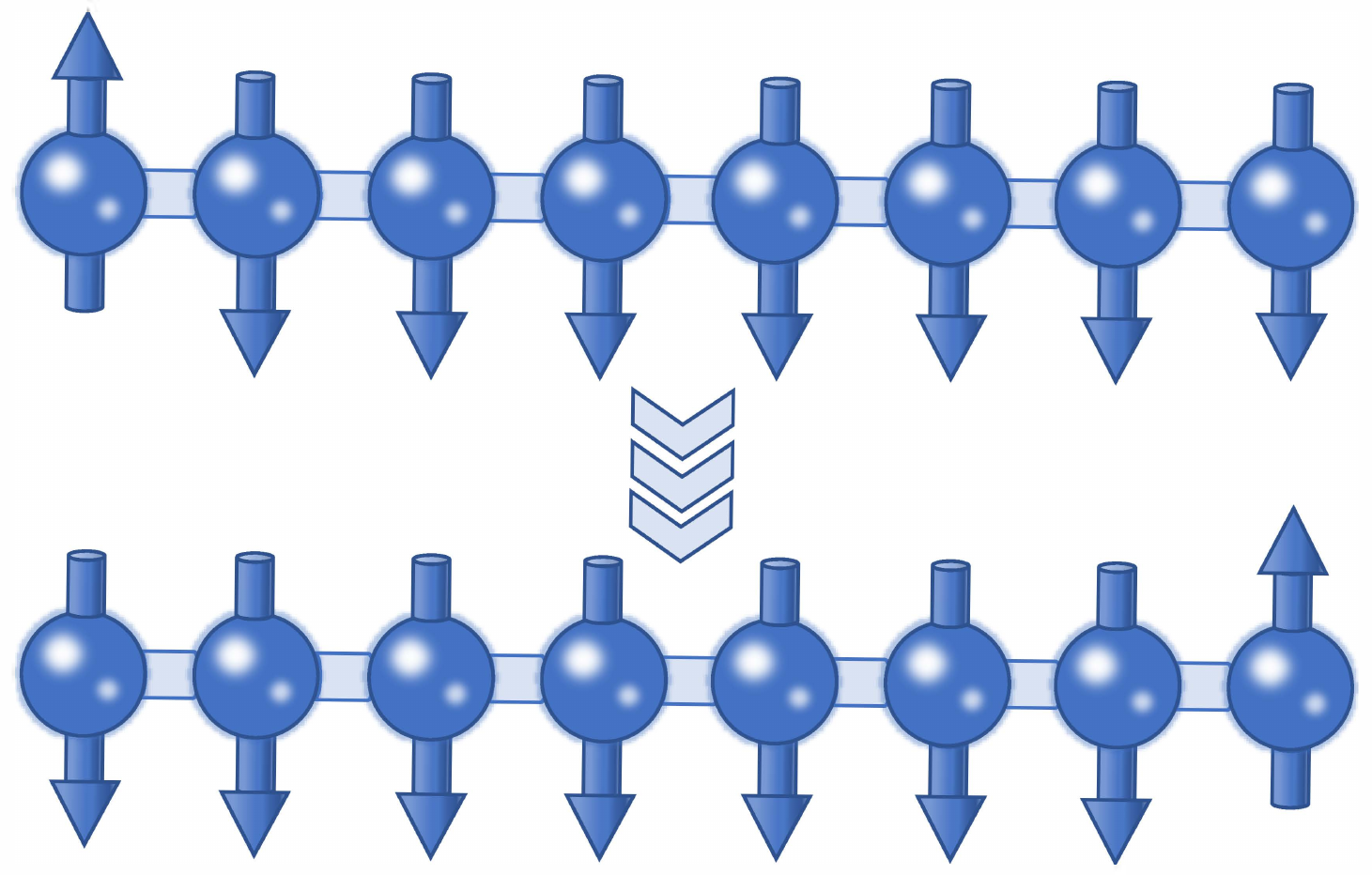}%
\label{fig_first_case}}
\caption{The 8-qubit spin-transfer system.}
\end{figure}

In the simulation, we set the total time as $T=(K-1)\pi/2$ and divide $T$ into $80$ equal time intervals.
\begin{figure*}[!t]
\centering
\subfloat[]{\includegraphics[width=3.5in]{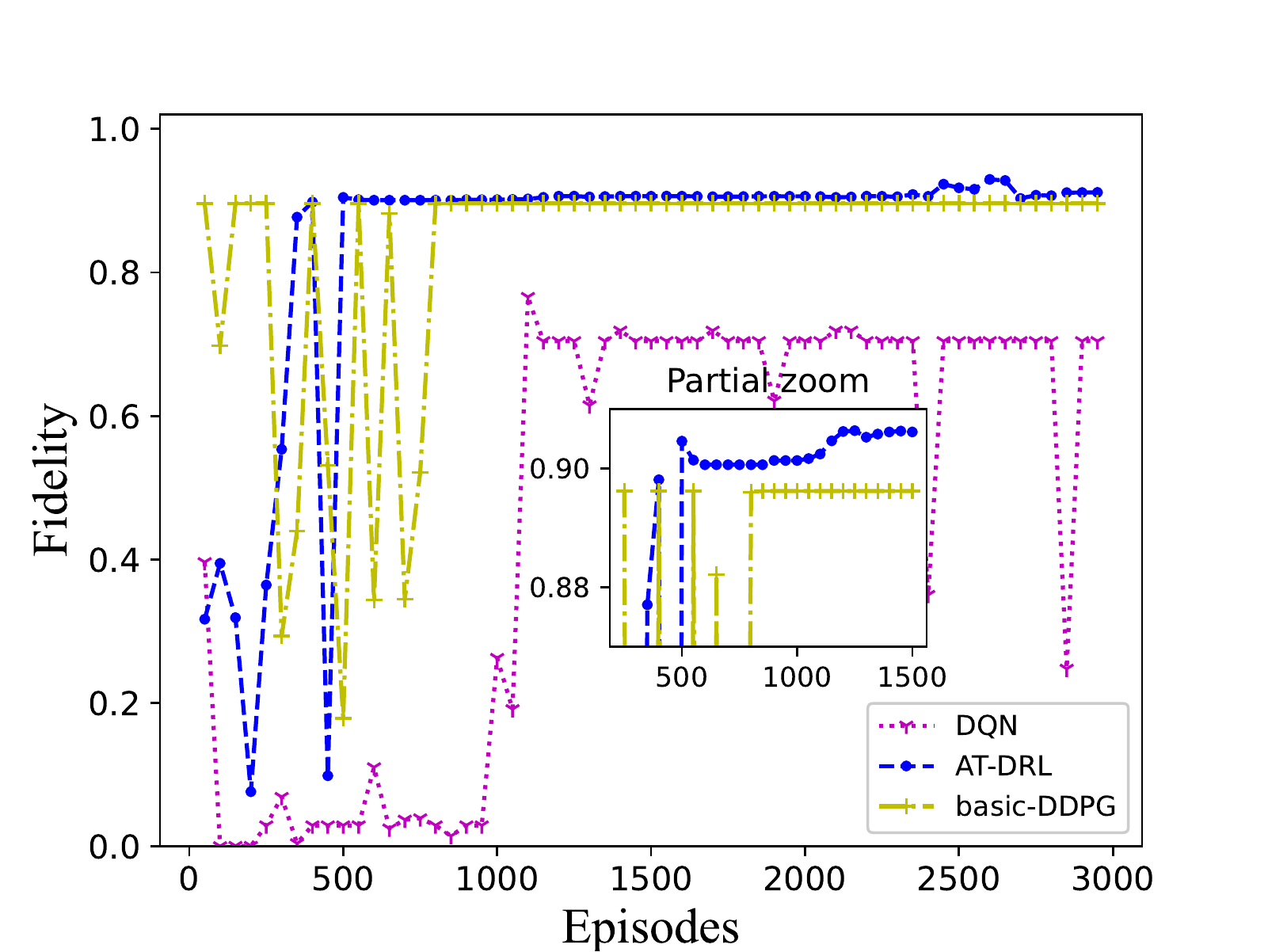}%
\label{fig_first_case}}
\hfil
\subfloat[]{\includegraphics[width=3.5in]{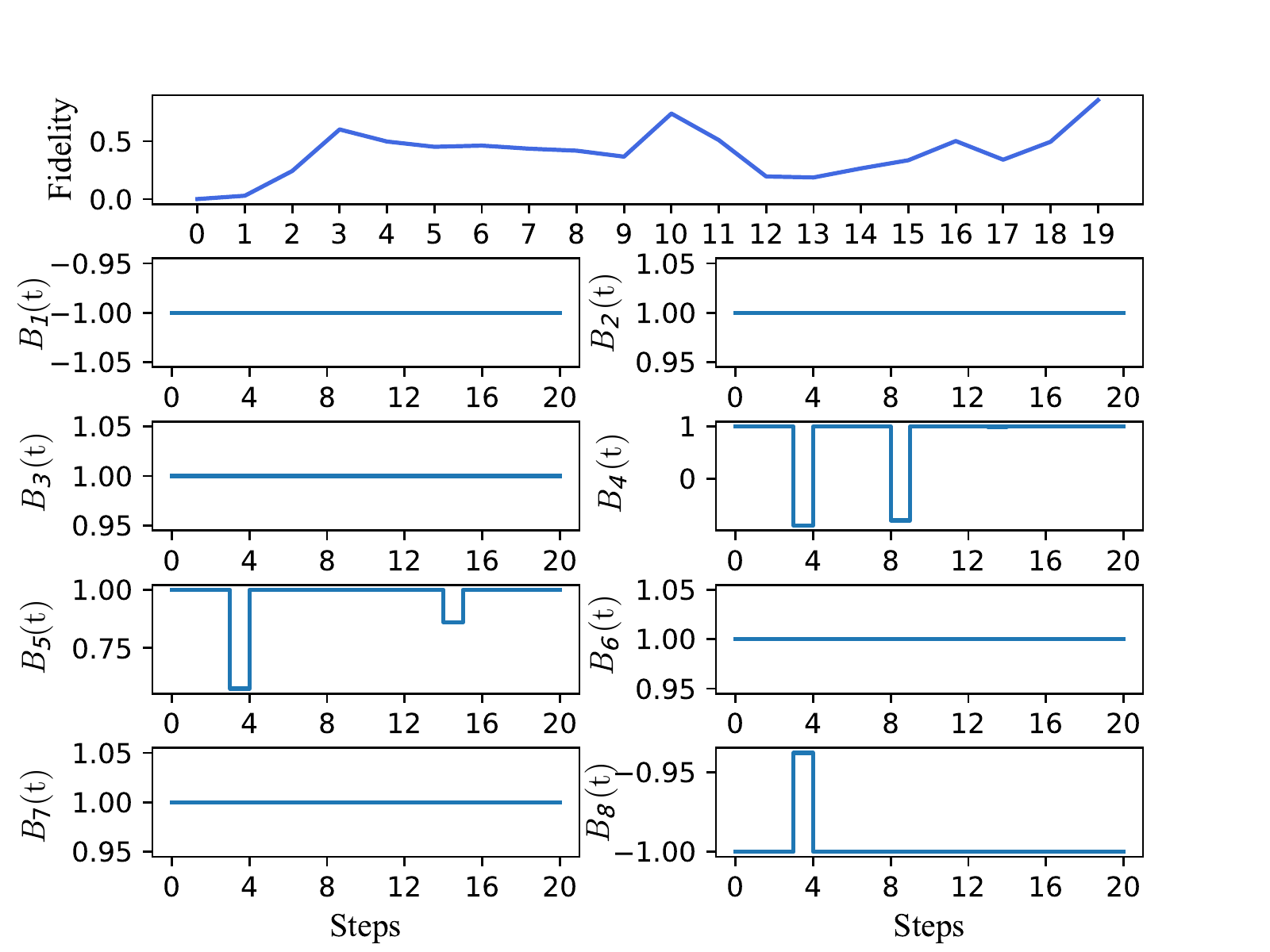}%
\label{fig_first_case}}
\caption{Quantum spin transfer of eight-qubit systems. (a) the whole evolution process and (b) the fidelity and the control laws of each step output of a complete episode.}
\label{fig_sim}
\end{figure*}
 The simulation results for eight-qubit state transfer is shown in Fig. 11(a), where the blue line achieves a fidelity of 0.90 at the 500-th episode. The yellow line indicates the basic DDPG, which has a slightly inferior result and reaches a fidelity of 0.894. In contrast, the DQN algorithm only attains a fidelity of 0.704. Fig. 11(b) shows that our algorithm finds a control strategy with 20 control pulses that eventually achieves a fidelity of 0.900. In the spin transfer task of the multi-qubit system, the network training process of the AT-DRL algorithm converges earlier and achieves better results in the end.

\section{Conclusion}

In this paper, we investigated the effectiveness of continuous control policies based on deep deterministic policy gradient and proposed an auxiliary task-based deep reinforcement learning (AT-DRL) method for quantum control. In particular, we designed a guided reward function based on the fidelity of quantum states and introduced an auxiliary task that shares training parameters with the main task to predict the reward provided. The numerical results of state preparation on different scenarios demonstrate the proposed method outperforms the basic DDPG algorithm in terms of fidelity, as well as greatly outperforms the basic DQN algorithm. In future work, we will extend it to open quantum systems to achieve accurate control even in the presence of environmental disturbances.

\vfill

\end{document}